\documentclass[aps,prb,reprint,amsmath,amssymb,superscriptaddress]
{revtex4-1}
\usepackage{graphicx}
\usepackage{color}
\usepackage[dvipdfm]{hyperref}
\hypersetup{
    colorlinks,
    linkcolor=blue,
    citecolor=blue,
    urlcolor=black,
}

\begin{document}

\title{Electronic and structural ground state of heavy alkali metals at high pressure}

\author{G.~Fabbris}
\altaffiliation[Current address: ]{Department of Condensed Matter Physics and Material Science, Brookhaven National Laboratory, Upton, NY 11973, USA }
\affiliation {Advanced Photon Source, Argonne National Laboratory, Argonne, IL 
60439, USA}
\affiliation {Department of Physics, Washington University, St. Louis, MO 
63130, USA}

\author{J.~Lim}
\affiliation {Department of Physics, Washington University, St. Louis, MO 
63130, USA}

\author{L.~S.~I.~Veiga}
\affiliation {Advanced Photon Source, Argonne National Laboratory, Argonne, IL 
60439, USA}
\affiliation {Laborat\'orio Nacional de Luz S\'incrotron, Campinas, SP, 13083-970, Brazil}
\affiliation {Instituto de F\'isica Gleb Wataghin, UNICAMP, Campinas, SP 
13083-859, Brazil}

\author{D.~Haskel}
\email[]{haskel@aps.anl.gov}
\affiliation {Advanced Photon Source, Argonne National Laboratory, Argonne, IL 
60439, USA}

\author{J.~S.~Schilling}
\email[]{jss@wuphys.wustl.edu}
\affiliation {Department of Physics, Washington University, St. Louis, MO 
63130, USA}

\date{\today}

\begin{abstract}
Alkali metals display unexpected properties at high pressure, including emergence of low symmetry crystal structures, that appear to occur due to enhanced electronic correlations among the otherwise nearly-free conduction electrons. We investigate the high pressure electronic and structural ground state of K, Rb, and Cs using x-ray absorption spectroscopy and x-ray diffraction measurements together with $ab$ $initio$ theoretical calculations. The sequence of phase transitions under pressure observed at low temperature is similar in all three heavy alkalis except for the absence of the $oC$84 phase in Cs. Both the experimental and theoretical results point to pressure-enhanced localization of the valence electrons characterized by pseudo-gap formation near the Fermi level and strong $spd$ hybridization. Although the crystal structures predicted to host magnetic order in K are not observed, the localization process appears to drive these alkalis closer to a strongly correlated electron state.
\end{abstract}

\maketitle

\section{Introduction}

The alkali metals were studied in the early days of quantum mechanics as the realization of a nearly free electron system.\cite{Wigner1933,Wigner1934} The weak interaction between its single $s$ valence electron and the heavily shielded atomic core leads to a very delocalized conduction band at ambient pressure, making these metals excellent electrical conductors, and favoring the high symmetry bcc crystal structure.\cite{Chi1979,McMahon2006a} Compression drastically changes this scenario, leading to highly unusual behavior such as metal-semiconductor-metal and metal-insulator transitions in Li and Na, respectively,\cite{Matsuoka2009, Matsuoka2014, Marques2011} enhanced resistivity in Rb and Cs,\cite{Stager1964, Jayaraman1967, McWhan1969, Wittig1970, Wittig1984} as well as superconductivity at relatively high temperatures in Li.\cite{Shimizu2002, Struzhkin2002, Deemyad2003} Additionally, pressure is believed to strongly enhance the $d$ character of the heavy alkalis (hereby defined as K, Rb and Cs) conduction band, enabling chemical reactions with transition metals.\cite{Parker1996} Finally, all alkalis display a bcc $\to$ fcc transition under pressure that is followed by remarkably low symmetry crystal structures.\cite{McMahon2006a} K and Rb assume an incommensurate host/guest (HG) structure at 19 and 16.6~GPa, respectively, \cite{Lundegaard2013,McMahon2001a} while Rb and Cs order in an orthorhombic phase with 52 and 84 atoms in the unit cell at 15 and 4.2~GPa, respectively.\cite{Nelmes2002, McMahon2001} The high pressure properties of alkali metals challenge the nearly free electron concept, exhibiting in manifold ways novel physics and chemistry.

The emergence of such low symmetry structures out of simple metals is a matter of great interest,\cite{Schwarz1998, Schwarz1999, Schwarz1999a,Hanfland2000, McMahon2001, McMahon2001a, Nelmes2002, Hanfland2002, Degtyareva2003, Ackland2004, Katzke2005, McMahon2006, Falconi2006, Marques2009, Lundegaard2009, Degtyareva2009, Pickard2011, Guillaume2011, Lundegaard2013} as it indicates that electronic interactions are relevant to the structural ground state. In a metal the electronic energy can be reduced by the introduction of a structural distortion that splits degenerate states at the Fermi level, lowering the overall energy.\cite{Jones1934,Mott1936} The driving mechanism for such distortions in alkalis is under debate.\cite{Neaton1999,Neaton2001,Ackland2004,Rousseau2008,Xie2007,Degtyareva2009,Pickard2011} Compression is argued to bring the impenetrable ionic cores together, localizing the valence electrons in the interstitial sites, hence reducing the bandwidth.\cite{Neaton1999, Neaton2001, Rousseau2008} Such reduced bandwidth favors a Peierls-like distortion, lowering the symmetry and electronic energy.\cite{Soderlind1995} This argument has been used to explain the low symmetry phases observed in some actinide metals\cite{Soderlind1995} whose structural behavior closely resembles the alkali series. \cite{Moore2009,Boehler1986,Gregoryanz2005,Guillaume2011,Schaeffer2012}  In fact, deviations from the nearly free electron behavior were observed in Na, K, and Rb.\cite{Loa2011} Furthermore, Fermi surface nesting is argued to be connected to a phonon mode softening that induces the bcc$\to$fcc transition in alkalis,\cite{Xie2000,Rodriguez-Prieto2006,Xie2008} and may signal the instability of the fcc phase at higher pressures.\cite{Xie2000,Xie2007}  Alternatively, these low symmetry structures have been suggested to follow the Hume-Rothery rules. \cite{Hume-Rothery1926,Ackland2004,Degtyareva2006,Degtyareva2009,Degtyareva2014} In this scenario the energy of the system is minimized by adopting a low symmetry phase in which the Brillouin zone (BZ) efficiently covers the nearly spherical Fermi surface (FS).\cite{Jones1934,Mott1936} Energy gaps open near the Fermi level by the FS-BZ interaction, reducing the electronic energy and generating a pseudo-gap in the density of states ($DOS$). The structures adopted by Hume-Rothery alloys depend only on the diameter of the FS, hence the number of valence electrons. Therefore, these models differ fundamentally on the nature of the valence electrons at high pressure: while the Peierls mechanism is driven by the localization of valence electrons, the Hume-Rothery mechanism is driven by a spherical ``nearly-free-electron-like'' Fermi surface.

Pickard and Needs have recently proposed that the electronic energy can be lowered by magnetic ordering.\cite{Pickard2011} The possible electronic localization enhances the density of states at the Fermi level ($DOS(E_F)$), potentially satisfying the Stoner criteria\cite{Stoner1939} for band magnetism ($DOS(E_F)I>1$, where $I$ is the exchange interaction). They predict that in Rb and Cs magnetic phases are close to stability at high pressure ($<$10~meV), and in K a ferromagnetic ground state occurs between $\approx$18.5-22~GPa. Interestingly, the predicted magnetic order occurs within lattice structures that are not observed in K at room temperature.\cite{McMahon2006, Marques2009,Lundegaard2013}
 Experimental investigation of the heavy alkalis at high-pressure has been focused on determining the complex crystal structure at room temperature, with few studies aimed at probing the electronic structure.\cite{Wittig1984,
Abd-Elmeguid1994,Tomita2005,Schilling2006,Loa2011} Furthermore, the extensive theoretical work\cite{Louie1974,McMahan1978,McMahan1984,Young1984,
Skriver1985,Ahuja2000,Xie2000,Schwarz2000,Ackland2004,Shi2006,Profeta2006,
Xie2007,Perez-Mato2007,Ma2008,Rousseau2008,Xie2008,Pickard2011,Degtyareva2014} has been mostly done at zero temperature. The additional challenge of introducing thermal fluctuations into calculations hampers the ability to understand the basic physical and chemical properties. 

In this work the high-pressure structural and electronic ground state of K, Rb and Cs is investigated using x-ray diffraction (XRD) and x-ray absorption near edge structure (XANES) at low temperature, as well as electronic structure calculations using both density functional theory (DFT) and real space multiple scattering approaches. None of the K, Rb or Cs crystal structures predicted to order magnetically were observed at 10~K, suggesting that magnetic order may not occur in these metals within the limits of this experiment. While for K the observed phase transitions reproduce those seen at room temperature, for Rb and Cs differences are seen in the phase boundaries. The Hume-Rothery mechanism\cite{Hume-Rothery1926,Jones1934,Mott1936} is inconsistent with the ground state structures of K and Cs, but cannot be completely discarded to drive the Rb-III phase. The orbital specific local $DOS$ ($LDOS$) indicates that pressure increases the $d$ level occupation through strong $spd$ hybridization. Therefore, both structural and electronic measurements give evidence that the electronic structure of the heavy alkali metals at high-pressure displays strong deviations from nearly-free-electron behavior.

\section{Experimental Details}

\subsection{Samples}

All experiments were performed using commercial samples (Sigma Aldrich - K, Cs 99.95\%, and Rb 99.6\%). The samples were shipped in vacuum ampoules which were broken inside an argon filled glove box where they were kept during the pressure cell loading; between experiments the samples were stored in a vacuum chamber. The alkalis are very soft and reactive, thus small pieces cut from the ingots were promptly loaded into the pressure cells. The absence of contaminants was verified by measuring powder diffraction in the sealed samples at room temperature before every experiment.

\subsection{X-ray diffraction}

High pressure XRD experiments were performed at the 16-BM-D (HPCAT) beamline of the Advanced Photon Source (APS), Argonne National Laboratory (ANL). Monochromatized x-rays (29.3~keV) were focused to 5x15~$\mu$m$^2$ using a pair of Kirkpatrick-Baez mirrors. Diffraction patterns were collected using a MAR345 image plate. Symmetric diamond anvil cells (DAC) (Princeton shops) were prepared with regular anvils of 600~$\mu$m and 300~$\mu$m culet diameter for Rb/Cs and K experiments, respectively. Rhenium gaskets were pre-indented to a thickness of $\sim$1/6 culet diameter. Boron carbide seats were used to increase the diffraction angular range (2$\theta _{max} \approx 25^{\circ}$). Ruby fluorescence was used to calibrate pressure.\cite{Chijioke2005} No pressure medium was used to prevent chemical reaction with the sample. No sign of reaction with the Re gasket or ruby was seen throughout the experiments. The DAC was kept at 10~K throughout the experiment using a He flow cryostat and pressure was applied $in$ $situ$ using a gearbox. The 2D patterns were converted into 1D plots using the Fit2D software.\cite{Hammersley1996} Strong texture was observed in every experiment. Consequently, unless otherwise specified, all XRD analyses were performed using the Le Bail method as implemented in the GSAS/EXPGUI program.\cite{Larson2000,Toby2001}

\subsection{X-ray absorption near edge structure}

High pressure XANES measurements were performed at K K-edge (3.608~keV), Rb K-edge (15.2~keV) and Cs L$_3$-edge (5.012~keV) at the 4-ID-D beamline of the APS, ANL. For Rb, a membrane driven CuBe DAC was prepared with a partially perforated diamond (100~$\mu$m wall) paired with a mini anvil (800~$\mu$m tall) glued on top of a fully perforated diamond.\cite{Dadashev2001} Ruby fluorescence was used to calibrate pressure.\cite{Chijioke2005} The same DAC was used for K and Cs, but the low energy of their absorption edges imposed the use of two partially perforated anvils. These perforations are opaque to visible light, thus pressure was determined by measuring the lattice parameter using the extended x-ray absorption fine structure (EXAFS) technique\cite{Haskel2007} and comparing to the equation of state derived by diffraction in this work (see below for further information). Diamonds with culet diameter of 300, 450 and 600~$\mu$m were used for K, Rb, and Cs, respectively. Rhenium gaskets were used for experiments on K, and stainless steel for Rb and Cs. For Rb, the gasket was pre-indented to 50~$\mu$m. Due to large sample absorption, the gaskets for K and Cs were pre-indented to 15~$\mu$m. The experiments were performed at 1.6~K using a He flow cryostat, and the temperature was increased to 15~K during pressure loading. A set of a Pd toroidal and Pd/Si flat mirrors was used to focus the x-rays to a spot of $\sim$150~$\mu$m diameter; the beam size was then further reduced to 50x50~$\mu$m$^2$ using slits. Harmonics were rejected using both the reflection cutoff of the mirrors and by detuning the monochromator. For the experiment on Rb, the intensity of the x-rays before and after the sample was measured using photodiodes, while for K and Cs, the incident intensity was measured with a He filled ion chamber, and the transmitted photons were detected with a photodiode placed inside the cryostat. XANES data was processed using the IFEFFIT/Horae package.\cite{Newville2001, Ravel2005}

\begin{figure}[t]
\includegraphics[width = 8.5 cm]{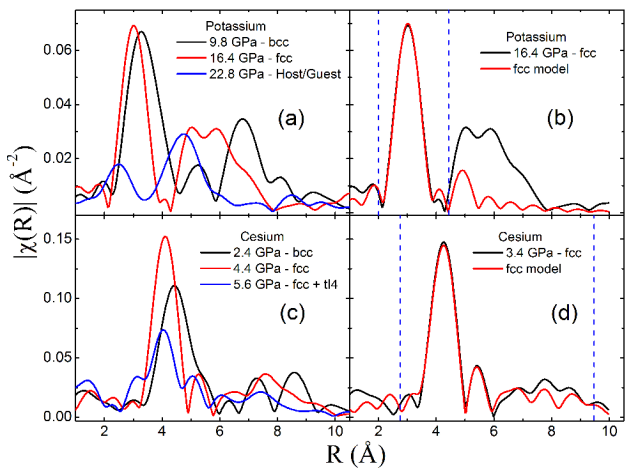} 
\caption{\label{exafs}(Color online) (a)(c) Pressure dependent EXAFS of K and Cs. Besides the strong reduction in the first coordination shell distance, changes in the second and third shells are indicative of the phase transitions. (b) Due to strong harmonic contamination, which primarily affects the EXAFS amplitude, only the first neighbors were fitted in K. (d) The high-quality Cs data was fitted to almost 10~\AA .}
\end{figure}

EXAFS is a well established technique for studying the local structure of both crystalline and amorphous samples.\cite{Sayers1971} Alkali metals are very soft (see Fig. \ref{diff2}), thus despite the limited accuracy of EXAFS for distance determination ($\approx$0.01~\AA), the large change of distances with pressure allows for reliable pressure calibration. Furthermore, the change in symmetry across the phase transitions is clearly seen in the data (Fig. \ref{exafs}), corroborating the obtained pressure. While Cs displays a rather symmetric $tI$4 structure at high-pressure, K-III is very complex. Thus for K, the pressures above 19~GPa were obtained from the linear relation between membrane and sample pressure.

\subsection{Electronic structure calculations}

XANES simulations were performed using the multiple scattering approach, in which the potentials are approximated as spherical muffin-tins, implemented in FEFF8.\cite{Ankudinov1998} This method describes the XANES as a superposition of scattering events connected by Green function propagators.\cite{Rehr2000} This formalism can also be used to calculate the electronic density, yielding orbital dependent $DOS$ that is used to interpret the data. Self-consistent potential calculations were performed using Hedin-Lundqvist self energy\cite{Hedin1970} in a cluster containing $\gtrsim$~100 atoms. XANES and $DOS$ were calculated using a $\gtrsim$~300 atoms cluster. DFT calculations using the WIEN2k code\cite{Blaha2001} were performed for Rb and Cs to verify the results obtained by FEFF8. A PBE-GGA exchange potential\cite{Perdew1996} was used with 10000 k-points for the bcc, fcc and Cs's $tI$4 structures, and 2000 k-points for Rb's $oC$52. Experimental lattice parameters were used in all calculations.

\section{Results}

\subsection{X-ray diffraction}

\begin{figure}[t]
\includegraphics[width = 7.5 cm]{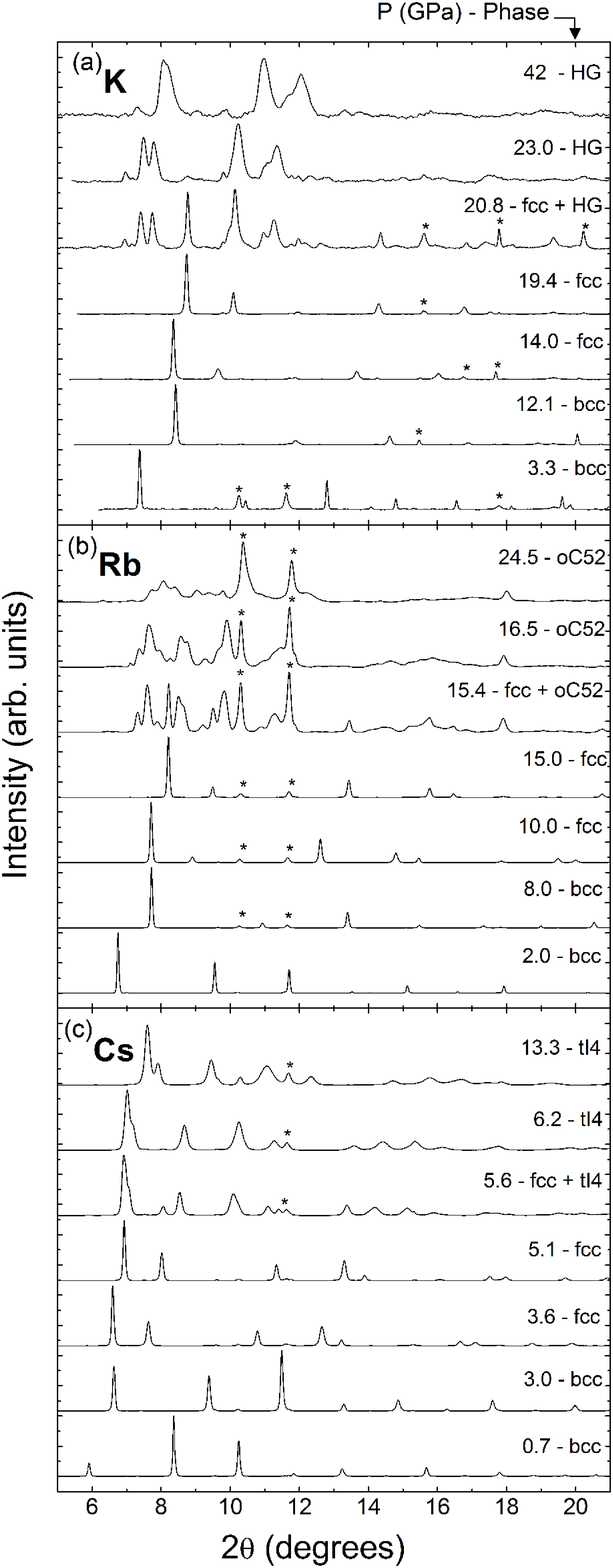} 
\caption{\label{diff1} Representative diffraction patterns for (a) K, (b) Rb, and (s) Cs. The phase transitions are clearly seen as the different symmetries lead to the appearance of new peaks. Bragg peaks from the rhenium gasket are marked with a $\ast$ symbol.}
\end{figure}

\paragraph{Potassium}

The phase transitions are clearly observed in the diffractogram by the appearance/suppression of peaks (Fig. \hyperref[diff1]{\ref{diff1}(a)}). The bcc to fcc transition occurs at 13$\pm$1~GPa, while the K-III phase becomes stable at 21$\pm$2~GPa. While no bcc/fcc coexistence was observed, at 20.8~GPa the fcc/K-III phases coexist. The high-symmetry bcc and fcc phases are easily indexed, but the unique determination of the post-fcc phase is more challenging. Most Bragg peaks observed above 21~GPa are consistent with the host structure ($I4/mcm$ space group) known to occur at room temperature.\cite{McMahon2006,Lundegaard2013} However, reflections from the associated guest phase ($C$-centered tetragonal) are expected to be very small, and only one reflection is visible at $2\theta \approx 8.7^{\circ}$. Attempts to index the diffractogram to other phases known to occur in the alkalis failed.

Magnetic order was predicted to occur in K between $\sim$18.5-22~GPa in crystal structures (simple cubic and c$I$16) that are not observed at room temperature.\cite{Pickard2011} These phases were not observed in this experiment. Nevertheless, if the predicted magnetic phases are ignored, DFT correctly finds K-III as the ground state above 20~GPa.\cite{Pickard2011}

\paragraph{Rubidium}

The bcc to fcc transition occurs at 8.9$\pm$1~GPa, at 15.7$\pm$1~GPa the Rb-III ($oC$52) phase is stable, with fcc/Rb-III coexistence seen at 15.7~GPa. Rb-III is stable to at least 24.5~GPa; at room temperature transitions to Rb-IV and Rb-V are observed in this pressure range (see Fig. \ref{phases}).

The $oC$52 structure ($C$222$_1$ unit cell) is remarkably complex. Its 52 unit cell atoms are distributed between seven inequivalent sites, and the structure refinement at room temperature was only possible through single crystal diffraction.\cite{Nelmes2002} In the Le Bail method no physical correlation is imposed on the diffraction intensities, thus the enormous number of reflections allowed by the $oC$52 phase complicates the determination of a unique structure. Therefore, the validity of this structure was verified by performing Rietveld refinements with atomic positions fixed to those found at room temperature.\cite{Nelmes2002} The data is reasonably well described by this refinement, suggesting that the $oC$52 is the correct crystal structure.

\paragraph{Cesium}
 
The bcc to fcc transition in Cs occurs at 3.4$\pm$0.3~GPa (Fig. \hyperref[diff1]{\ref{diff1}(c)}). Further pressure leads to fcc/Cs-IV ($tI$4) coexistence between 5.4$\pm$0.1~GPa and 6.1$\pm$0.3~GPa, after which only the Cs-IV is observed to at least 13.4~GPa. The very low symmetry Cs-III ($oC$84), stable only between 4.2~GPa and 4.3~GPa at room temperature,\cite{McMahon2001} is not seen in the present data. Even though the pressure step ($\approx$0.3~GPa) prevents a definite answer, the observed fcc-$tI$4 coexistence at 5.5 and 5.8~GPa is evidence that the Cs-III phase does not occur at low temperature. The same conclusion was reached through DFT.\cite{Pickard2011}

\begin{figure}[t]
\includegraphics[width = 8.5 cm]{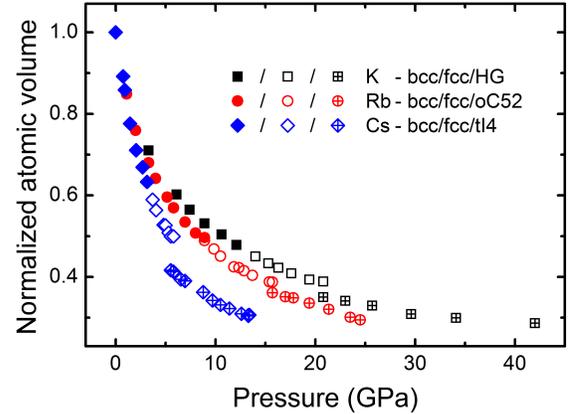} 
\caption{\label{diff2} (Color online) Atomic volume as a function of pressure. The following ambient pressure low temperature atomic volume was used to normalize the data: K - 71.507~\AA $^3$, Rb - 87.338~\AA $^3$, Cs - 110.617~\AA $^3$.}
\end{figure}

\begin{table}[b]
\caption{\label{eos} Atomic volume (V$_0$), bulk modulus (B$_0$), and pressure derivative of the bulk modulus (B$_0 ^{\prime}$) as obtained here and from the literature. DFT-FP stands for density functional theory using the full-potential linear augmented plane-wave method.}
\begin{ruledtabular}
\begin{tabular}{c c c c c c}
Alkali & Method & $T$ (K) & $V_0$ (\AA$^3$) & $B_0$ (GPa) & $B_0 ^{\prime}$ \\
\hline
K  & this work & 10 & 75(2) & 4.2(9) & 3.5(1) \\
   & piston-diplacement\cite{Anderson1983} & 4 & 75.7 & 3.7 & 4.1 \\
   & x-ray diffraction\cite{Winzenick1994} & 300 & 75.65 & 2.96 & 4.06 \\
   & DFT-FP\cite{Xie2008} & 0 & 74.09 & 3.68 & 3.66 \\ \\
Rb & this work & 10 & 95(3) & 3.1(6) & 3.5(1) \\
   & piston-diplacement\cite{Anderson1983} & 4 & 92.6 & 2.9 & 4.1 \\
   & x-ray diffraction\cite{Winzenick1994} & 300 & 92.74 & 2.3 & 4.1 \\
   & DFT-FP\cite{Xie2008} & 0 & 93.07 & 2.84 & 3.52 \\ \\
Cs & this work & 10 & 118(1) & 2.8(5) & 3.1(1) \\
   & piston-diplacement\cite{Anderson1985} & 4 & 110.4 & 2.1 & 4.0 \\
   & DFT-FP\cite{Xie2008} & 0 & 112.68 & 2.29 & 3.17 \\
\end{tabular}
\end{ruledtabular}
\end{table}

\begin{figure*}[t]
\includegraphics[width = 17.1 cm]{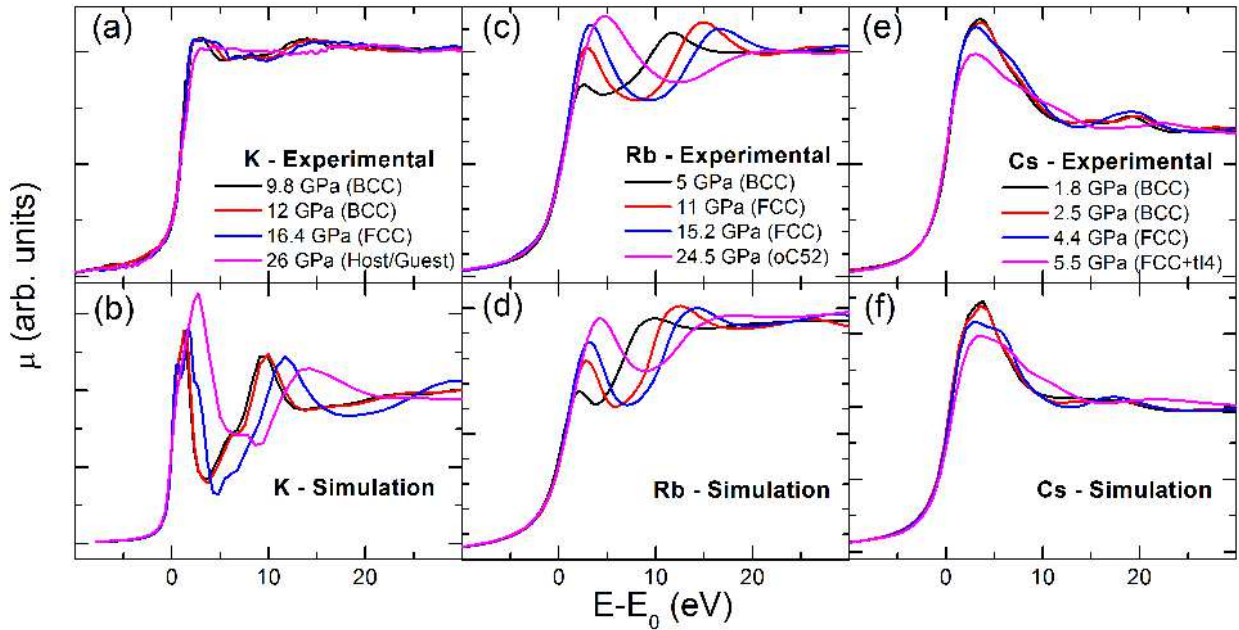} 
\caption{\label{xanes}(Color online) Panels (a), (c), and (e) display the pressure dependence of XANES data for K, Rb, and Cs, respectively. The FEFF simulations are shown below the raw data on panels (b), (d), and (f). Experimental data was shifted in energy by less than 1 eV for better comparison.}
\end{figure*}

\paragraph{Equation of state at low temperature}

The equation of state (EOS) of K, Rb, and Cs obtained at 10~K is shown in Fig. \ref{diff2}. These were fit to a 3$^{rd}$ order Birch-Murnaghan EOS\cite{Birch1947} up to the fcc$\to$post-fcc transition (Table \ref{eos}). The results are consistent with previous measurements. Differences in the bulk modulus pressure derivative ($B_0 ^{\prime}$) obtained here and through piston-displacement method are likely due to the much reduced pressure range in that experiment ($\sim$2~GPa).\cite{Anderson1983,Anderson1985} Furthermore, the smaller bulk modulus ($B_0$) seen in previous diffraction experiments\cite{Winzenick1994} is consistent with the different temperatures.

A large volume discontinuity (volume collapse) is observed across the fcc$\to$post-fcc transition in all heavy alkalis, reaching 3.9\% (K), 2.6\% (Rb), and 8.6\% (Cs) (volume collapse size defined as $(V_{fcc}-V_{post-fcc})/V_{0}$). Volume collapse transitions have been observed in many elemental solids (e.g. Refs. \onlinecite{McMahan1998,Lindbaum2003,Akahama2005,Samudrala2012,Fabbris2013}, and references therein), being typically argued to signal the onset of changes in electronic properties, such as 4$f$ bonding and/or Kondo effect in lanthanides.\cite{Johansson1974,Allen1982,Rueff2006,Bradley2012,
Lipp2012,Fabbris2013,Johansson2014}

\subsection{X-ray absorption near edge structure}

The measured x-ray absorption spectra at the K K-edge (3.608~keV) are affected by harmonic contamination (Fig. \hyperref[xanes]{\ref{xanes}(a)}). Attempts to improve data quality by detuning the monochromator and by using the reflectivity cutoff of the Pd/Si mirrors were unsuccessful. Despite not allowing a meaningful comparison with calculated spectra, the very small shift of the absorption edge to 40~GPa ($\lesssim$0.5 eV) clearly demonstrate the absence of a valence increase in K this pressure. This shows that the proposed 3$p$-conduction band mixing does not occur at the fcc$\to$HG transition.\cite{Degtyareva2014}

\begin{figure}[b]
\includegraphics[width =8.5 cm]{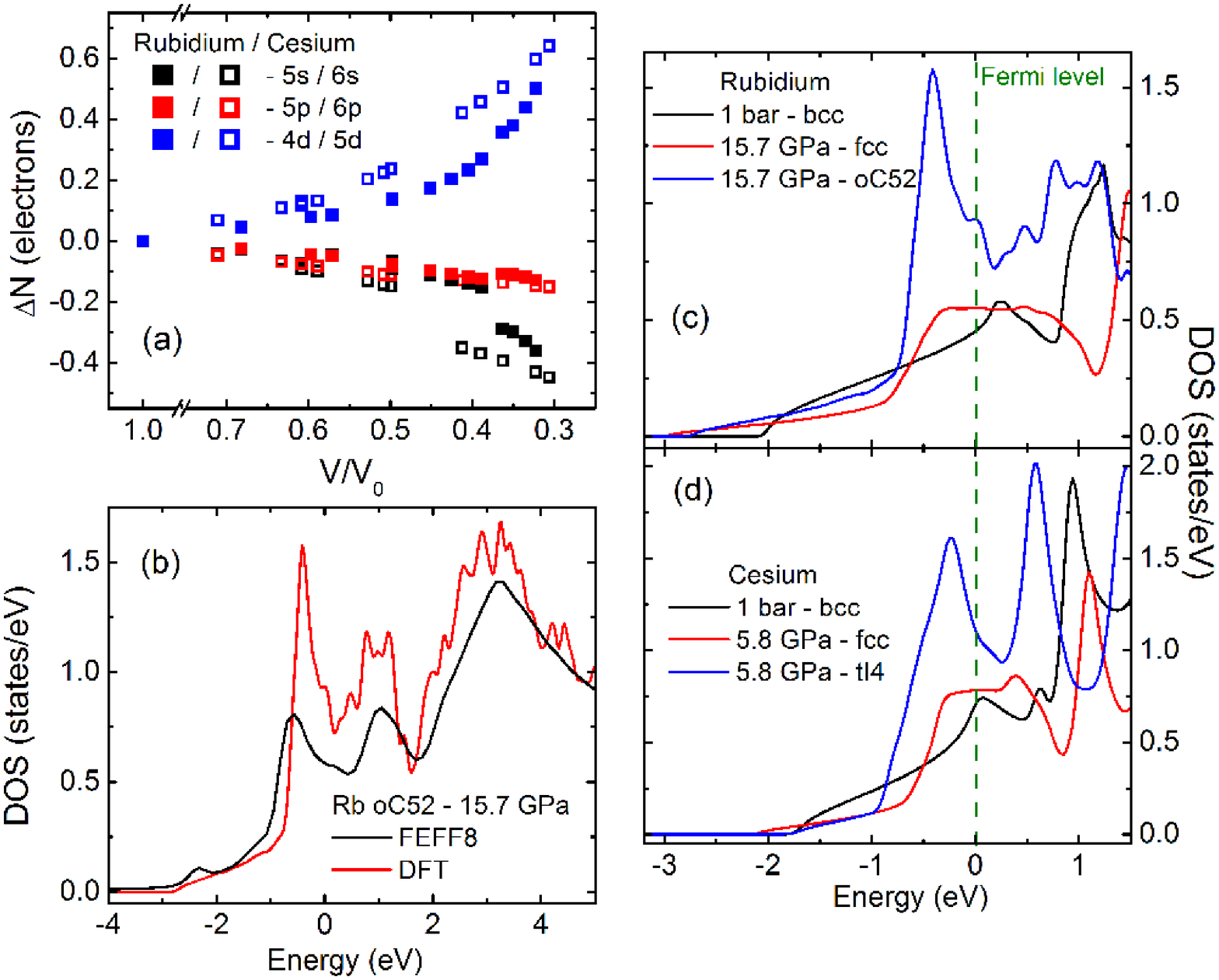} 
\caption{\label{ecount}(Color online) (a) Variation of orbital occupation with pressure for Rb and Cs. A break was introduced in the horizontal axis for better data display. (b) Similar Rb-III $DOS$ calculated by FEFF8 and DFT provides further evidence for the proper description of the high pressure electronic structure. (c-d) Pressure dependence of Rb's and Cs's $DOS$ calculated by DFT. The low symmetry phase opens a pseudo-gap at the Fermi level, partially localizing the valence electrons.}
\end{figure}

\begin{figure*}[t]
\includegraphics[width = 18 cm]{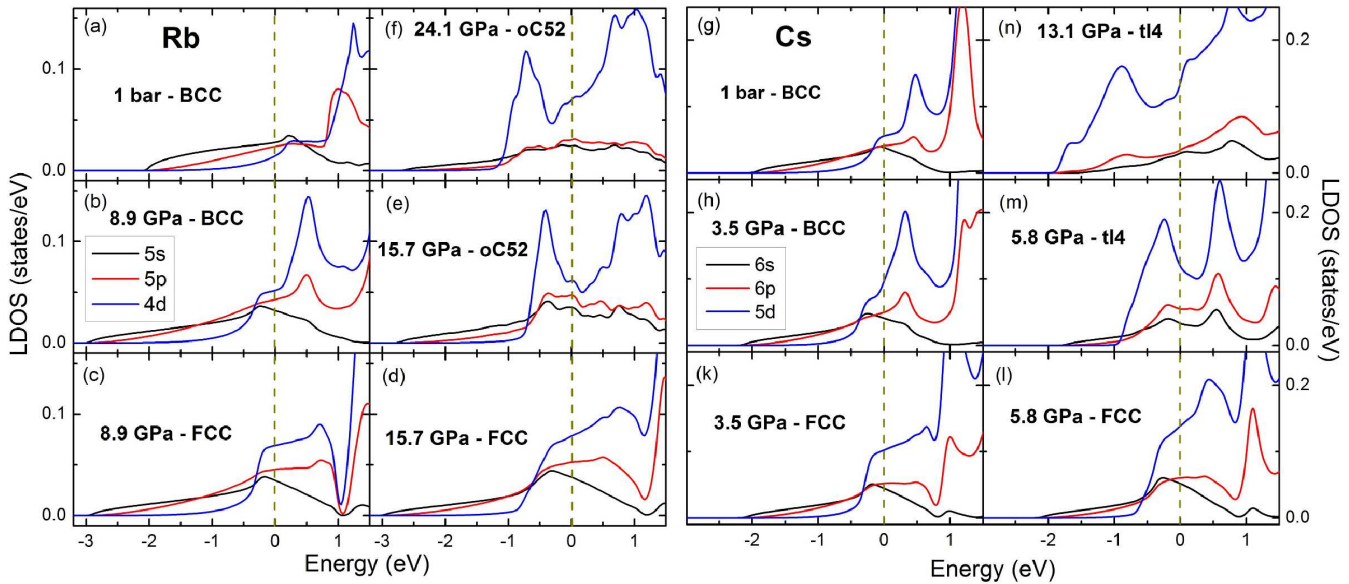} 
\caption{\label{dos}(Color online) LDOS for (a-f) Rb, and (g-n) Cs, as a function of pressure.}
\end{figure*}

The excelent quality of the Rb K-edge and Cs L$_3$-edge allows a direct comparison to the calculated spectra (Fig. \hyperref[xanes]{\ref{xanes}(c-f)}). The absorption cross section is dominated by the dipolar transition, the K-edge being sensitive to the density of empty $p$ states, while the L$_3$ edge to the empty $d$ states. Pressure induces substantial changes in the Rb and Cs XANES data. In the former, the strong increase in the lowest energy peak indicates an increased number of empty $p$ states, whereas in the later the first peak is suppressed, pointing to a reduction in the empty $d$ states. Therefore, the data for Rb and Cs are qualitatively consistent with an enhanced $d$ occupation at the cost of $sp$ electrons.

\subsection{Electronic structure}

The agreement between experiment and simulation validates the calculated electronic structure. The number of electrons per orbital calculated by FEFF8 is displayed in Fig. \hyperref[ecount]{\ref{ecount}(a)} as a function of relative atomic volume (V/V$_0$). Both Rb and Cs display enhanced $d$ character in the conduction band at high pressures. Noticeably, little increase in $d$ level occupation occurs up to the bcc$\to$fcc transitions (V/V$_0$ $\approx$ 0.5 and 0.6 for Rb and Cs, respectively). Furthermore, a nearly continuous change in $d$ occupation is observed across the large volume collapse observed at the fcc$\to$post-fcc transition.
The emergence of the low symmetry phases marks the onset of a hastening of electron transfer from $s$ to $d$ states. The phase transitions in Rb and Cs correlate with the number of $s$ electrons (N$_s$). The bcc$\to$fcc transition occurs at N$_s$ = 0.54$\pm$0.01 electrons, and the fcc$\to$post-fcc at N$_s$ = 0.46$\pm$0.01 electrons. Such remarkable similarity in s orbital occupation suggests that the deformation of the Fermi surface away from the low pressure spherical shape plays a significant role in both transitions.

While the $DOS$ calculated by FEFF8 is a reasonable approximation (Fig. \hyperref[ecount]{\ref{ecount}(b)}), the broader features observed in the $DOS$ are likely due to the overlapping muffin-tin approximation,\cite{Mattheiss1964} which treats the interatomic potential as constant. For Rb and Cs pressure leads to strong $spd$ hybridization observed in the $LDOS$ for each orbital (Fig. \ref{dos}). Across the fcc$\to$post-fcc transition, a clear splitting of the $DOS$ around the Fermi level is observed (Fig. \hyperref[ecount]{\ref{ecount}(c)}). This is consistent with the emergence of low symmetry phases through minimization of the electronic energy by the opening of a pseudo-gap. Furthermore, an abrupt localization of valence electrons is suggested by the sharper features observed in the occupied $DOS$ across the fcc$\to$post-fcc transition.

\section{Discussion}

\subsection{Temperature dependence of crystal structure}

\begin{figure}[t]
\includegraphics[width = 8.5 cm]{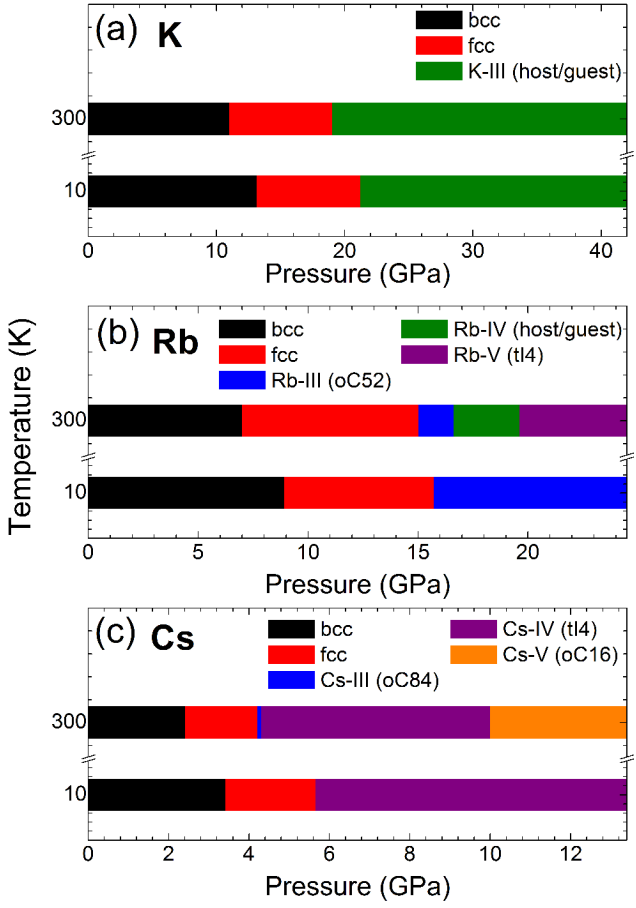} 
\caption{\label{phases}(Color online) Phases of K, Rb, and Cs at 10~K and room temperature. \cite{McMahon2006a,Lundegaard2013}}
\end{figure}

While K displays the same phases at low and room temperature to 42~GPa, Rb and Cs display relevant differences in their phase diagrams (Fig. \ref{phases}). In Rb, the o$C$52 phase is stable to at least 24.5~GPa, overcoming the range where HG and t$I$4 are stable at room temperature. In Cs, the low symmetry o$C$84 phase is not observed. The o$C$52 and o$C$84 phases exhibit the same type of layered structure, displaying different layer order.\cite{Nelmes2002} The mismatch between these layers has been argued to be unstable to sliding, which would explain their short range of stability ($\approx$0.1~GPa in Cs, \cite{McMahon2001} and $\approx$1.6~GPa in Rb \cite{Nelmes2002} at room temperature).\cite{Perez-Mato2007} While the extended o$C$52 stability in Rb at low temperature supports this argument, the absence of the o$C$84 in Cs appears to contradict it. However, DFT calculations indicate that the o$C$84 simply becomes energetically unstable at low temperature.\cite{Pickard2011}

\subsection{Emergence of low symmetry crystal structures in simple metals}

Pressure-induced enhanced $d$ character of valence electrons has been argued to drive the stability of low symmetry phases due to the anisotropy of the $d$ wavefunction.\cite{Sternheimer1950, Louie1974, McMahan1978, McMahon2001, Nelmes2002} In fact, 6$s\to$5$d$ charge transfer has been suggested by M$\rm{\ddot{o}}$ssbauer data to occur in Cs.\cite{Abd-Elmeguid1994} However, in a rigid band approximation, simple $s \to d$ charge transfer leads to a sequence of close-packed crystal structures, such as those observed in the lanthanides.\cite{Duthie1977,Skriver1985} Furthermore, S$\rm{\ddot{o}}$derlind $et$ $al.$ have shown that metallic bonding is insensitive to the anisotropy of the valence electron wavefunctions.\cite{Soderlind1995}

\begin{figure}[b]
\includegraphics[width = 8.5 cm]{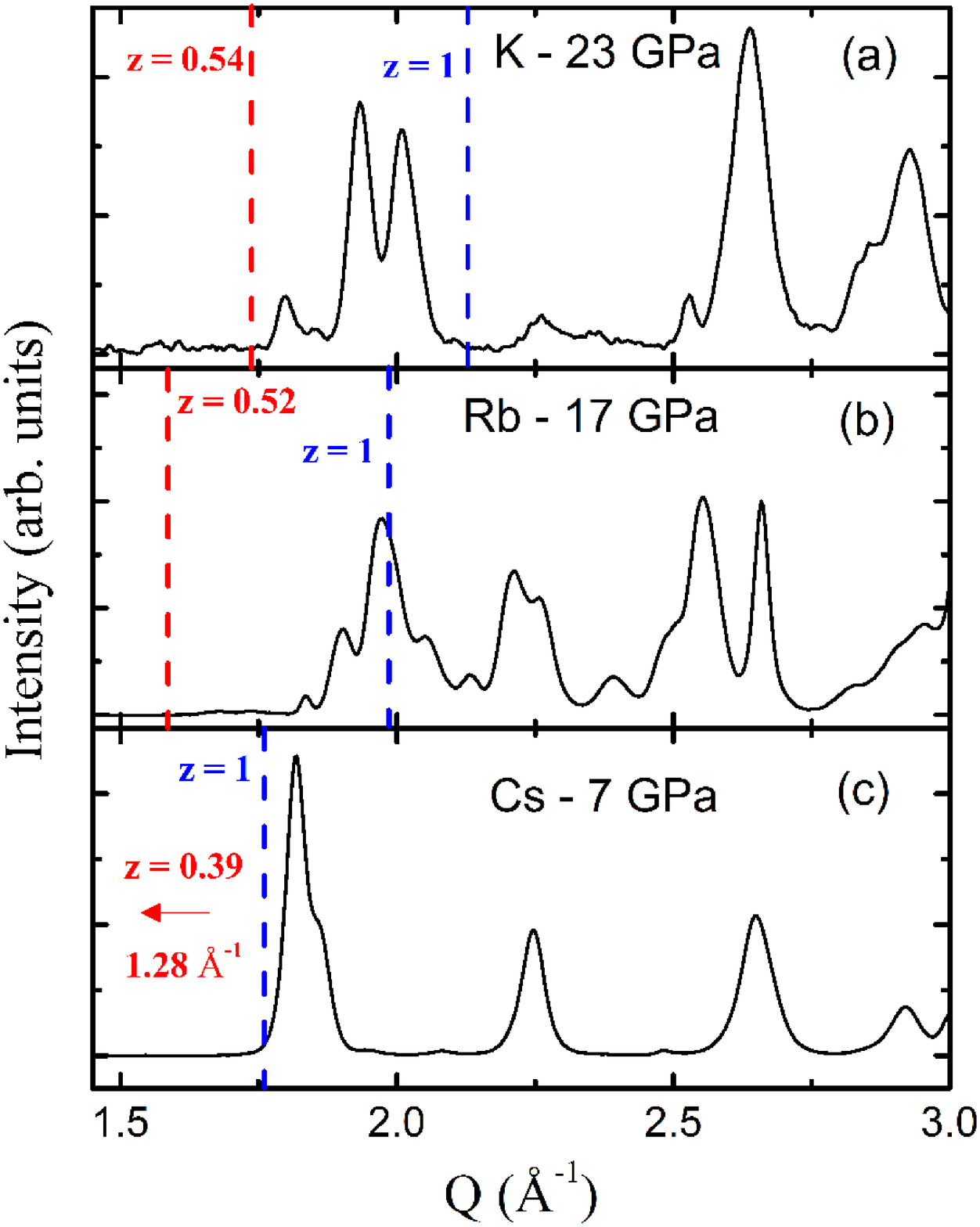} 
\caption{\label{fsbz}(Color online) Diffractogram of post-fcc phases for K (a), Rb (b), and Cs (c) compared to the Fermi sphere radius for z = 1 and z as calculated by FEFF8.}
\end{figure}

Perhaps the most prevalent explanation for the high pressure phase transitions observed in the alkalis is the Hume-Rothery mechanism through the FS-BZ interaction.\cite{Ackland2004,Degtyareva2003,Degtyareva2006,Degtyareva2014} This mechanism can be experimentally verified by two properties: the ``closeness" factor ($\eta$ = 2k$_F$/q, where k$_F$ is the radius of the Fermi surface - assumed to be spherical - and q is the reciprocal lattice vector of a Bragg reflection) and the volume of the BZ occupied by the FS (V$_{FS}$/V$_{BZ}$). For 1.0 $< \eta < $1.05, the FS lies very close to the BZ boundary, opening a pseudo-gap that reduces the overall electronic energy. The number of valence electrons used to calculate k$_F$ is usually taken as the number of $sp$ electrons (z) only, as $d$ states tend to strongly deform the spherical FS.\cite{Ackland2004} In Fig. \ref{fsbz} the diffractograms of K, Rb, and Cs within their post-fcc phases are compared to k$_F$ calculated assuming a purely spherical FS (z = 1) and using the $sp$ occupation obtained by FEFF8. Only Rb in the z = 1 case appears to be consistent with the Hume-Rothery mechanism. Furthermore, using the strongest Bragg peaks nearby k$_F$ for z = 1 in K and Cs, V$_{FS}$/V$_{BZ}$ is 113\% and 75\%, respectively, while for Rb this ratio is 94.3\%. Although the diffraction data is consistent with the o$C$52 phase of Rb being stabilized by a Hume-Rothery mechanism, the remarkable similarity between the experimental and simulated XANES is strong evidence that z = 0.52 (not z = 1) is correct, which would lead to a strongly distorted FS and $\eta$ = 0.81. Additionally, it would be difficult to propose different mechanisms for these alkalis given the similarity between their $DOS$ across the fcc$\to$post-fcc transition (Fig. \ref{dos}). It has been proposed that the HG structure of K occurs due to an increased number of valence electrons (2.6 \={e}/atom) arising from hybridization with the inner 3$p$ level.\cite{Degtyareva2014} This is in strong disagreement with the present experimental and calculated results (even at 40~GPa, the 3$p$ level lies more than 12~eV below the Fermi level).

The sequence of structure transitions across the actinide series is remarkably similar to that in alkali metals under pressure, including the emergence of very low symmetry phases together with much reduced melting temperatures.\cite{Moore2009,Boehler1986,Gregoryanz2005,Guillaume2011,Schaeffer2012} It has been argued that the structure of metals can be explained by a competition between the Madelung energy, favoring high symmetry phases, and the Peierls distortion, favoring low symmetry structures.\cite{Soderlind1995} Pressure weakens the nearly free electron behavior of valence electrons by introducing sharp features in the $DOS$ that indicate electronic localization (Fig. \ref{dos}). At high pressure, $spd$ hybridization leads to a dominating $d$ character in the valence electrons, bringing the heavy alkalis closer to transition metals. In fact, Rb and Cs display features about 1.5~eV wide across the Fermi level in the fcc phase, in excellent agreement with the predicted $\approx$1-2 eV bandwidth necessary for a Peierls distortion to occur in Fe.\cite{Ackland2004} Therefore, the present results indicate that the low symmetry structures of Rb and Cs, and likely K, occur due to pressure induced localization of the conduction band, within a Peierls mechanism.\cite{Soderlind1995, Neaton1999}

\subsection{Magnetic order in heavy alkalis at high pressure}

The emergence of magnetic order in heavy alkali metals arising from high-pressure electronic localization is an exciting possibility.\cite{Pickard2011} The observed crystal structures at 10~K suggest absence of magnetic ordering in these metals as predicted using DFT (Fig. \ref{diff1}). However, only magnetic measurements will be able to definitely address this question. Furthermore, the multiple nearly degenerate phases ($\lesssim$~10~meV) found by DFT suggest that strain effects present in the current non-hydrostatic measurements may contribute to this result. Even in the absence of magnetic ordering, the pressure-induced localization of valence electrons should lead to a larger paramagnetic response, which would also be of interest. Nevertheless, if the predicted magnetic crystal structures for K are ignored, DFT correctly predicts the ground state post-fcc phases for K, Rb and Cs.\cite{Pickard2011}

\section{Conclusion} 

In this paper the electronic and structural ground state of the heavy alkalis was investigated. These metals were shown to transition to low symmetry phases at similar pressures at 10 and 300~K. Understanding how such low symmetry phases occur in simple metals is of general interest. Here, a combination of x-ray diffraction and spectroscopy measurements, as well as theoretical calculations, provide a unique insight into the origin of such phases. It is shown that pressure partially localizes the conduction band of K, Rb, and Cs, distorting the otherwise nearly-free-electron-like valence band. This process evolves through strong $spd$ hybridization. The localization of valence electrons is argued to trigger a Peierls transition, in which electronic energy is gained by opening a pseudo-gap at the Fermi level. Compression is demonstrated to push the valence band towards the strongly correlated electron regime, likely triggering similar behavior to that observed in transition  metals. 

\begin{acknowledgments}
Work at Argonne National Laboratory is supported by the U.S. Department of Energy, Office of Science, Office of Basic Energy Sciences, under Contract No. DE-AC02-06CH11357. This work was supported by the National Science Foundation (NSF) through Grant No. DMR-1104742 and by the Carnegie/DOE Alliance Center (CDAC) through NNSA/DOE Grant No. DE-FC52-08NA28554. L. S. I. Veiga was supported by FAPESP (SP-Brazil) under Contract No. 2013/14338-3. HPCAT operations are supported by DOE-NNSA under Award No. DE-NA0001974 and DOE-BES under Award No. DE-FG02-99ER45775, with partial instrumentation funding by NSF. The authors would like to thank Curtis Kenney-Benson, Changyong Park, and Dmitry Popov for their support in the x-ray diffraction measurements.
\end{acknowledgments}

\bibliography{alkali_ref}

\begin{thebibliography}{100}%
\makeatletter
\providecommand \@ifxundefined [1]{%
 \@ifx{#1\undefined}
}%
\providecommand \@ifnum [1]{%
 \ifnum #1\expandafter \@firstoftwo
 \else \expandafter \@secondoftwo
 \fi
}%
\providecommand \@ifx [1]{%
 \ifx #1\expandafter \@firstoftwo
 \else \expandafter \@secondoftwo
 \fi
}%
\providecommand \natexlab [1]{#1}%
\providecommand \enquote  [1]{``#1''}%
\providecommand \bibnamefont  [1]{#1}%
\providecommand \bibfnamefont [1]{#1}%
\providecommand \citenamefont [1]{#1}%
\providecommand \href@noop [0]{\@secondoftwo}%
\providecommand \href [0]{\begingroup \@sanitize@url \@href}%
\providecommand \@href[1]{\@@startlink{#1}\@@href}%
\providecommand \@@href[1]{\endgroup#1\@@endlink}%
\providecommand \@sanitize@url [0]{\catcode `\\12\catcode `\$12\catcode
  `\&12\catcode `\#12\catcode `\^12\catcode `\_12\catcode `\%12\relax}%
\providecommand \@@startlink[1]{}%
\providecommand \@@endlink[0]{}%
\providecommand \url  [0]{\begingroup\@sanitize@url \@url }%
\providecommand \@url [1]{\endgroup\@href {#1}{\urlprefix }}%
\providecommand \urlprefix  [0]{URL }%
\providecommand \Eprint [0]{\href }%
\providecommand \doibase [0]{http://dx.doi.org/}%
\providecommand \selectlanguage [0]{\@gobble}%
\providecommand \bibinfo  [0]{\@secondoftwo}%
\providecommand \bibfield  [0]{\@secondoftwo}%
\providecommand \translation [1]{[#1]}%
\providecommand \BibitemOpen [0]{}%
\providecommand \bibitemStop [0]{}%
\providecommand \bibitemNoStop [0]{.\EOS\space}%
\providecommand \EOS [0]{\spacefactor3000\relax}%
\providecommand \BibitemShut  [1]{\csname bibitem#1\endcsname}%
\let\auto@bib@innerbib\@empty

\hypersetup{urlcolor=blue}

\bibitem [{\citenamefont {Wigner}\ and\ \citenamefont
  {Seitz}(1933)}]{Wigner1933}%
  \BibitemOpen
  \bibfield  {author} {\bibinfo {author} {\bibfnamefont {E.}~\bibnamefont
  {Wigner}}\ and\ \bibinfo {author} {\bibfnamefont {F.}~\bibnamefont {Seitz}},\
  }\href {\doibase 10.1103/PhysRev.43.804} {\bibfield  {journal} {\bibinfo
  {journal} {Phys. Rev.}\ }\textbf {\bibinfo {volume} {43}},\ \bibinfo {pages}
  {804} (\bibinfo {year} {1933})}\BibitemShut {NoStop}%
\bibitem [{\citenamefont {Wigner}\ and\ \citenamefont
  {Seitz}(1934)}]{Wigner1934}%
  \BibitemOpen
  \bibfield  {author} {\bibinfo {author} {\bibfnamefont {E.}~\bibnamefont
  {Wigner}}\ and\ \bibinfo {author} {\bibfnamefont {F.}~\bibnamefont {Seitz}},\
  }\href {\doibase 10.1103/PhysRev.46.509} {\bibfield  {journal} {\bibinfo
  {journal} {Phys. Rev.}\ }\textbf {\bibinfo {volume} {46}},\ \bibinfo {pages}
  {509} (\bibinfo {year} {1934})}\BibitemShut {NoStop}%
\bibitem [{\citenamefont {Chi}(1979)}]{Chi1979}%
  \BibitemOpen
  \bibfield  {author} {\bibinfo {author} {\bibfnamefont {T.~C.}\ \bibnamefont
  {Chi}},\ }\href {\doibase 10.1063/1.555598} {\bibfield  {journal} {\bibinfo
  {journal} {J. Phys. Chem. Ref. Data}\ }\textbf {\bibinfo {volume} {8}},\
  \bibinfo {pages} {339} (\bibinfo {year} {1979})},\ \bibinfo {note} {and
  references therein}\BibitemShut {NoStop}%
\bibitem [{\citenamefont {McMahon}\ and\ \citenamefont
  {Nelmes}(2006)}]{McMahon2006a}%
  \BibitemOpen
  \bibfield  {author} {\bibinfo {author} {\bibfnamefont {M.~I.}\ \bibnamefont
  {McMahon}}\ and\ \bibinfo {author} {\bibfnamefont {R.~J.}\ \bibnamefont
  {Nelmes}},\ }\href {\doibase 10.1039/b517777b} {\bibfield  {journal}
  {\bibinfo  {journal} {Chem. Soc. Rev.}\ }\textbf {\bibinfo {volume} {35}},\
  \bibinfo {pages} {943} (\bibinfo {year} {2006})},\ \bibinfo {note} {and
  references therein}\BibitemShut {NoStop}%
\bibitem [{\citenamefont {Matsuoka}\ and\ \citenamefont
  {Shimizu}(2009)}]{Matsuoka2009}%
  \BibitemOpen
  \bibfield  {author} {\bibinfo {author} {\bibfnamefont {T.}~\bibnamefont
  {Matsuoka}}\ and\ \bibinfo {author} {\bibfnamefont {K.}~\bibnamefont
  {Shimizu}},\ }\href {\doibase 10.1038/nature07827} {\bibfield  {journal}
  {\bibinfo  {journal} {Nature}\ }\textbf {\bibinfo {volume} {458}},\ \bibinfo
  {pages} {186} (\bibinfo {year} {2009})}\BibitemShut {NoStop}%
\bibitem [{\citenamefont {Matsuoka}\ \emph {et~al.}(2014)\citenamefont
  {Matsuoka}, \citenamefont {Sakata}, \citenamefont {Nakamoto}, \citenamefont
  {Takahama}, \citenamefont {Ichimaru}, \citenamefont {Mukai}, \citenamefont
  {Ohta}, \citenamefont {Hirao}, \citenamefont {Ohishi},\ and\ \citenamefont
  {Shimizu}}]{Matsuoka2014}%
  \BibitemOpen
  \bibfield  {author} {\bibinfo {author} {\bibfnamefont {T.}~\bibnamefont
  {Matsuoka}}, \bibinfo {author} {\bibfnamefont {M.}~\bibnamefont {Sakata}},
  \bibinfo {author} {\bibfnamefont {Y.}~\bibnamefont {Nakamoto}}, \bibinfo
  {author} {\bibfnamefont {K.}~\bibnamefont {Takahama}}, \bibinfo {author}
  {\bibfnamefont {K.}~\bibnamefont {Ichimaru}}, \bibinfo {author}
  {\bibfnamefont {K.}~\bibnamefont {Mukai}}, \bibinfo {author} {\bibfnamefont
  {K.}~\bibnamefont {Ohta}}, \bibinfo {author} {\bibfnamefont {N.}~\bibnamefont
  {Hirao}}, \bibinfo {author} {\bibfnamefont {Y.}~\bibnamefont {Ohishi}}, \
  and\ \bibinfo {author} {\bibfnamefont {K.}~\bibnamefont {Shimizu}},\ }\href
  {\doibase 10.1103/PhysRevB.89.144103} {\bibfield  {journal} {\bibinfo
  {journal} {Phys. Rev. B}\ }\textbf {\bibinfo {volume} {89}},\ \bibinfo
  {pages} {144103} (\bibinfo {year} {2014})}\BibitemShut {NoStop}%
\bibitem [{\citenamefont {Marqu\'{e}s}\ \emph {et~al.}(2011)\citenamefont
  {Marqu\'{e}s}, \citenamefont {McMahon}, \citenamefont {Gregoryanz},
  \citenamefont {Hanfland}, \citenamefont {Guillaume}, \citenamefont {Pickard},
  \citenamefont {Ackland},\ and\ \citenamefont {Nelmes}}]{Marques2011}%
  \BibitemOpen
  \bibfield  {author} {\bibinfo {author} {\bibfnamefont {M.}~\bibnamefont
  {Marqu\'{e}s}}, \bibinfo {author} {\bibfnamefont {M.}~\bibnamefont
  {McMahon}}, \bibinfo {author} {\bibfnamefont {E.}~\bibnamefont {Gregoryanz}},
  \bibinfo {author} {\bibfnamefont {M.}~\bibnamefont {Hanfland}}, \bibinfo
  {author} {\bibfnamefont {C.}~\bibnamefont {Guillaume}}, \bibinfo {author}
  {\bibfnamefont {C.}~\bibnamefont {Pickard}}, \bibinfo {author} {\bibfnamefont
  {G.}~\bibnamefont {Ackland}}, \ and\ \bibinfo {author} {\bibfnamefont
  {R.}~\bibnamefont {Nelmes}},\ }\href {\doibase
  10.1103/PhysRevLett.106.095502} {\bibfield  {journal} {\bibinfo  {journal}
  {Phys. Rev. Lett.}\ }\textbf {\bibinfo {volume} {106}},\ \bibinfo {pages} {4}
  (\bibinfo {year} {2011})}\BibitemShut {NoStop}%
\bibitem [{\citenamefont {Stager}\ and\ \citenamefont
  {Drickamer}(1964)}]{Stager1964}%
  \BibitemOpen
  \bibfield  {author} {\bibinfo {author} {\bibfnamefont {R.}~\bibnamefont
  {Stager}}\ and\ \bibinfo {author} {\bibfnamefont {H.}~\bibnamefont
  {Drickamer}},\ }\href {\doibase 10.1103/PhysRevLett.12.19} {\bibfield
  {journal} {\bibinfo  {journal} {Phys. Rev. Lett.}\ }\textbf {\bibinfo
  {volume} {12}},\ \bibinfo {pages} {19} (\bibinfo {year} {1964})}\BibitemShut
  {NoStop}%
\bibitem [{\citenamefont {Jayaraman}\ \emph {et~al.}(1967)\citenamefont
  {Jayaraman}, \citenamefont {Newton},\ and\ \citenamefont
  {McDonough}}]{Jayaraman1967}%
  \BibitemOpen
  \bibfield  {author} {\bibinfo {author} {\bibfnamefont {A.}~\bibnamefont
  {Jayaraman}}, \bibinfo {author} {\bibfnamefont {R.}~\bibnamefont {Newton}}, \
  and\ \bibinfo {author} {\bibfnamefont {J.}~\bibnamefont {McDonough}},\ }\href
  {\doibase 10.1103/PhysRev.159.527} {\bibfield  {journal} {\bibinfo  {journal}
  {Phys. Rev.}\ }\textbf {\bibinfo {volume} {159}},\ \bibinfo {pages} {527}
  (\bibinfo {year} {1967})}\BibitemShut {NoStop}%
\bibitem [{\citenamefont {McWhan}\ and\ \citenamefont
  {Stevens}(1969)}]{McWhan1969}%
  \BibitemOpen
  \bibfield  {author} {\bibinfo {author} {\bibfnamefont {D.}~\bibnamefont
  {McWhan}}\ and\ \bibinfo {author} {\bibfnamefont {A.}~\bibnamefont
  {Stevens}},\ }\href {\doibase 10.1016/0038-1098(69)90405-0} {\bibfield
  {journal} {\bibinfo  {journal} {Solid State Commun.}\ }\textbf {\bibinfo
  {volume} {7}},\ \bibinfo {pages} {301} (\bibinfo {year} {1969})}\BibitemShut
  {NoStop}%
\bibitem [{\citenamefont {Wittig}(1970)}]{Wittig1970}%
  \BibitemOpen
  \bibfield  {author} {\bibinfo {author} {\bibfnamefont {J.}~\bibnamefont
  {Wittig}},\ }\href {\doibase 10.1103/PhysRevLett.24.812} {\bibfield
  {journal} {\bibinfo  {journal} {Phys. Rev. Lett.}\ }\textbf {\bibinfo
  {volume} {24}},\ \bibinfo {pages} {812} (\bibinfo {year} {1970})}\BibitemShut
  {NoStop}%
\bibitem [{\citenamefont {Wittig}(1984)}]{Wittig1984}%
  \BibitemOpen
  \bibfield  {author} {\bibinfo {author} {\bibfnamefont {J.}~\bibnamefont
  {Wittig}},\ }\href@noop {} {\bibfield  {journal} {\bibinfo  {journal} {Mat.
  Res. Soc. Symp. Proc.}\ }\textbf {\bibinfo {volume} {22}},\ \bibinfo {pages}
  {17} (\bibinfo {year} {1984})}\BibitemShut {NoStop}%
\bibitem [{\citenamefont {Shimizu}\ \emph {et~al.}(2002)\citenamefont
  {Shimizu}, \citenamefont {Ishikawa}, \citenamefont {Takao}, \citenamefont
  {Yagi},\ and\ \citenamefont {Amaya}}]{Shimizu2002}%
  \BibitemOpen
  \bibfield  {author} {\bibinfo {author} {\bibfnamefont {K.}~\bibnamefont
  {Shimizu}}, \bibinfo {author} {\bibfnamefont {H.}~\bibnamefont {Ishikawa}},
  \bibinfo {author} {\bibfnamefont {D.}~\bibnamefont {Takao}}, \bibinfo
  {author} {\bibfnamefont {T.}~\bibnamefont {Yagi}}, \ and\ \bibinfo {author}
  {\bibfnamefont {K.}~\bibnamefont {Amaya}},\ }\href {\doibase
  10.1038/nature01098} {\bibfield  {journal} {\bibinfo  {journal} {Nature}\
  }\textbf {\bibinfo {volume} {419}},\ \bibinfo {pages} {597} (\bibinfo {year}
  {2002})}\BibitemShut {NoStop}%
\bibitem [{\citenamefont {Struzhkin}\ \emph {et~al.}(2002)\citenamefont
  {Struzhkin}, \citenamefont {Eremets}, \citenamefont {Gan}, \citenamefont
  {Mao},\ and\ \citenamefont {Hemley}}]{Struzhkin2002}%
  \BibitemOpen
  \bibfield  {author} {\bibinfo {author} {\bibfnamefont {V.~V.}\ \bibnamefont
  {Struzhkin}}, \bibinfo {author} {\bibfnamefont {M.~I.}\ \bibnamefont
  {Eremets}}, \bibinfo {author} {\bibfnamefont {W.}~\bibnamefont {Gan}},
  \bibinfo {author} {\bibfnamefont {H.-K.}\ \bibnamefont {Mao}}, \ and\
  \bibinfo {author} {\bibfnamefont {R.~J.}\ \bibnamefont {Hemley}},\ }\href
  {\doibase 10.1126/science.1078535} {\bibfield  {journal} {\bibinfo  {journal}
  {Science}\ }\textbf {\bibinfo {volume} {298}},\ \bibinfo {pages} {1213}
  (\bibinfo {year} {2002})}\BibitemShut {NoStop}%
\bibitem [{\citenamefont {Deemyad}\ and\ \citenamefont
  {Schilling}(2003)}]{Deemyad2003}%
  \BibitemOpen
  \bibfield  {author} {\bibinfo {author} {\bibfnamefont {S.}~\bibnamefont
  {Deemyad}}\ and\ \bibinfo {author} {\bibfnamefont {J.~S.}\ \bibnamefont
  {Schilling}},\ }\href {\doibase 10.1103/PhysRevLett.91.167001} {\bibfield
  {journal} {\bibinfo  {journal} {Phys. Rev. Lett.}\ }\textbf {\bibinfo
  {volume} {91}},\ \bibinfo {pages} {14} (\bibinfo {year} {2003})}\BibitemShut
  {NoStop}%
\bibitem [{\citenamefont {Parker}\ \emph {et~al.}(1996)\citenamefont {Parker},
  \citenamefont {Atou},\ and\ \citenamefont {Badding}}]{Parker1996}%
  \BibitemOpen
  \bibfield  {author} {\bibinfo {author} {\bibfnamefont {L.~J.}\ \bibnamefont
  {Parker}}, \bibinfo {author} {\bibfnamefont {T.}~\bibnamefont {Atou}}, \ and\
  \bibinfo {author} {\bibfnamefont {J.~V.}\ \bibnamefont {Badding}},\ }\href
  {\doibase 10.1126/science.273.5271.95} {\bibfield  {journal} {\bibinfo
  {journal} {Science}\ }\textbf {\bibinfo {volume} {273}},\ \bibinfo {pages}
  {95} (\bibinfo {year} {1996})}\BibitemShut {NoStop}%
\bibitem [{\citenamefont {Lundegaard}\ \emph {et~al.}(2013)\citenamefont
  {Lundegaard}, \citenamefont {Stinton}, \citenamefont {Zelazny}, \citenamefont
  {Guillaume}, \citenamefont {Proctor}, \citenamefont {Loa}, \citenamefont
  {Gregoryanz}, \citenamefont {Nelmes},\ and\ \citenamefont
  {McMahon}}]{Lundegaard2013}%
  \BibitemOpen
  \bibfield  {author} {\bibinfo {author} {\bibfnamefont {L.~F.}\ \bibnamefont
  {Lundegaard}}, \bibinfo {author} {\bibfnamefont {G.~W.}\ \bibnamefont
  {Stinton}}, \bibinfo {author} {\bibfnamefont {M.}~\bibnamefont {Zelazny}},
  \bibinfo {author} {\bibfnamefont {C.~L.}\ \bibnamefont {Guillaume}}, \bibinfo
  {author} {\bibfnamefont {J.~E.}\ \bibnamefont {Proctor}}, \bibinfo {author}
  {\bibfnamefont {I.}~\bibnamefont {Loa}}, \bibinfo {author} {\bibfnamefont
  {E.}~\bibnamefont {Gregoryanz}}, \bibinfo {author} {\bibfnamefont {R.~J.}\
  \bibnamefont {Nelmes}}, \ and\ \bibinfo {author} {\bibfnamefont {M.~I.}\
  \bibnamefont {McMahon}},\ }\href {\doibase 10.1103/PhysRevB.88.054106}
  {\bibfield  {journal} {\bibinfo  {journal} {Phys. Rev. B}\ }\textbf {\bibinfo
  {volume} {88}},\ \bibinfo {pages} {054106} (\bibinfo {year}
  {2013})}\BibitemShut {NoStop}%
\bibitem [{\citenamefont {McMahon}\ \emph
  {et~al.}(2001{\natexlab{a}})\citenamefont {McMahon}, \citenamefont {Rekhi},\
  and\ \citenamefont {Nelmes}}]{McMahon2001a}%
  \BibitemOpen
  \bibfield  {author} {\bibinfo {author} {\bibfnamefont {M.}~\bibnamefont
  {McMahon}}, \bibinfo {author} {\bibfnamefont {S.}~\bibnamefont {Rekhi}}, \
  and\ \bibinfo {author} {\bibfnamefont {R.}~\bibnamefont {Nelmes}},\ }\href
  {\doibase 10.1103/PhysRevLett.87.055501} {\bibfield  {journal} {\bibinfo
  {journal} {Phys. Rev. Lett.}\ }\textbf {\bibinfo {volume} {87}},\ \bibinfo
  {pages} {055501} (\bibinfo {year} {2001}{\natexlab{a}})}\BibitemShut
  {NoStop}%
\bibitem [{\citenamefont {Nelmes}\ \emph {et~al.}(2002)\citenamefont {Nelmes},
  \citenamefont {McMahon}, \citenamefont {Loveday},\ and\ \citenamefont
  {Rekhi}}]{Nelmes2002}%
  \BibitemOpen
  \bibfield  {author} {\bibinfo {author} {\bibfnamefont {R.}~\bibnamefont
  {Nelmes}}, \bibinfo {author} {\bibfnamefont {M.}~\bibnamefont {McMahon}},
  \bibinfo {author} {\bibfnamefont {J.}~\bibnamefont {Loveday}}, \ and\
  \bibinfo {author} {\bibfnamefont {S.}~\bibnamefont {Rekhi}},\ }\href
  {\doibase 10.1103/PhysRevLett.88.155503} {\bibfield  {journal} {\bibinfo
  {journal} {Phys. Rev. Lett.}\ }\textbf {\bibinfo {volume} {88}},\ \bibinfo
  {pages} {155503} (\bibinfo {year} {2002})}\BibitemShut {NoStop}%
\bibitem [{\citenamefont {McMahon}\ \emph
  {et~al.}(2001{\natexlab{b}})\citenamefont {McMahon}, \citenamefont {Nelmes},\
  and\ \citenamefont {Rekhi}}]{McMahon2001}%
  \BibitemOpen
  \bibfield  {author} {\bibinfo {author} {\bibfnamefont {M.}~\bibnamefont
  {McMahon}}, \bibinfo {author} {\bibfnamefont {R.}~\bibnamefont {Nelmes}}, \
  and\ \bibinfo {author} {\bibfnamefont {S.}~\bibnamefont {Rekhi}},\ }\href
  {\doibase 10.1103/PhysRevLett.87.255502} {\bibfield  {journal} {\bibinfo
  {journal} {Phys. Rev. Lett.}\ }\textbf {\bibinfo {volume} {87}},\ \bibinfo
  {pages} {255502} (\bibinfo {year} {2001}{\natexlab{b}})}\BibitemShut
  {NoStop}%
\bibitem [{\citenamefont {Schwarz}\ \emph {et~al.}(1998)\citenamefont
  {Schwarz}, \citenamefont {Takemura}, \citenamefont {Hanfland},\ and\
  \citenamefont {Syassen}}]{Schwarz1998}%
  \BibitemOpen
  \bibfield  {author} {\bibinfo {author} {\bibfnamefont {U.}~\bibnamefont
  {Schwarz}}, \bibinfo {author} {\bibfnamefont {K.}~\bibnamefont {Takemura}},
  \bibinfo {author} {\bibfnamefont {M.}~\bibnamefont {Hanfland}}, \ and\
  \bibinfo {author} {\bibfnamefont {K.}~\bibnamefont {Syassen}},\ }\href
  {\doibase 10.1103/PhysRevLett.81.2711} {\bibfield  {journal} {\bibinfo
  {journal} {Phys. Rev. Lett.}\ }\textbf {\bibinfo {volume} {81}},\ \bibinfo
  {pages} {2711} (\bibinfo {year} {1998})}\BibitemShut {NoStop}%
\bibitem [{\citenamefont {Schwarz}\ \emph
  {et~al.}(1999{\natexlab{a}})\citenamefont {Schwarz}, \citenamefont
  {Grzechnik}, \citenamefont {Syassen}, \citenamefont {Loa},\ and\
  \citenamefont {Hanfland}}]{Schwarz1999}%
  \BibitemOpen
  \bibfield  {author} {\bibinfo {author} {\bibfnamefont {U.}~\bibnamefont
  {Schwarz}}, \bibinfo {author} {\bibfnamefont {a.}~\bibnamefont {Grzechnik}},
  \bibinfo {author} {\bibfnamefont {K.}~\bibnamefont {Syassen}}, \bibinfo
  {author} {\bibfnamefont {I.}~\bibnamefont {Loa}}, \ and\ \bibinfo {author}
  {\bibfnamefont {M.}~\bibnamefont {Hanfland}},\ }\href {\doibase
  10.1103/PhysRevLett.83.4085} {\bibfield  {journal} {\bibinfo  {journal}
  {Phys. Rev. Lett.}\ }\textbf {\bibinfo {volume} {83}},\ \bibinfo {pages}
  {4085} (\bibinfo {year} {1999}{\natexlab{a}})}\BibitemShut {NoStop}%
\bibitem [{\citenamefont {Schwarz}\ \emph
  {et~al.}(1999{\natexlab{b}})\citenamefont {Schwarz}, \citenamefont {Syassen},
  \citenamefont {Grzechnik},\ and\ \citenamefont {Hanfland}}]{Schwarz1999a}%
  \BibitemOpen
  \bibfield  {author} {\bibinfo {author} {\bibfnamefont {U.}~\bibnamefont
  {Schwarz}}, \bibinfo {author} {\bibfnamefont {K.}~\bibnamefont {Syassen}},
  \bibinfo {author} {\bibfnamefont {A.}~\bibnamefont {Grzechnik}}, \ and\
  \bibinfo {author} {\bibfnamefont {M.}~\bibnamefont {Hanfland}},\ }\href
  {\doibase 10.1016/S0038-1098(99)00362-2} {\bibfield  {journal} {\bibinfo
  {journal} {Solid State Commun.}\ }\textbf {\bibinfo {volume} {112}},\
  \bibinfo {pages} {319} (\bibinfo {year} {1999}{\natexlab{b}})}\BibitemShut
  {NoStop}%
\bibitem [{\citenamefont {Hanfland}\ \emph {et~al.}(2000)\citenamefont
  {Hanfland}, \citenamefont {Syassen}, \citenamefont {Christensen},\ and\
  \citenamefont {Novikov}}]{Hanfland2000}%
  \BibitemOpen
  \bibfield  {author} {\bibinfo {author} {\bibfnamefont {M.}~\bibnamefont
  {Hanfland}}, \bibinfo {author} {\bibfnamefont {K.}~\bibnamefont {Syassen}},
  \bibinfo {author} {\bibfnamefont {N.~E.}\ \bibnamefont {Christensen}}, \ and\
  \bibinfo {author} {\bibfnamefont {D.~L.}\ \bibnamefont {Novikov}},\ }\href
  {\doibase 10.1038/35041515} {\bibfield  {journal} {\bibinfo  {journal}
  {Nature}\ }\textbf {\bibinfo {volume} {408}},\ \bibinfo {pages} {174}
  (\bibinfo {year} {2000})}\BibitemShut {NoStop}%
\bibitem [{\citenamefont {Hanfland}\ \emph {et~al.}(2002)\citenamefont
  {Hanfland}, \citenamefont {Loa},\ and\ \citenamefont
  {Syassen}}]{Hanfland2002}%
  \BibitemOpen
  \bibfield  {author} {\bibinfo {author} {\bibfnamefont {M.}~\bibnamefont
  {Hanfland}}, \bibinfo {author} {\bibfnamefont {I.}~\bibnamefont {Loa}}, \
  and\ \bibinfo {author} {\bibfnamefont {K.}~\bibnamefont {Syassen}},\ }\href
  {\doibase 10.1103/PhysRevB.65.184109} {\bibfield  {journal} {\bibinfo
  {journal} {Phys. Rev. B}\ }\textbf {\bibinfo {volume} {65}},\ \bibinfo
  {pages} {184109} (\bibinfo {year} {2002})}\BibitemShut {NoStop}%
\bibitem [{\citenamefont {Degtyareva}(2003)}]{Degtyareva2003}%
  \BibitemOpen
  \bibfield  {author} {\bibinfo {author} {\bibfnamefont {V.~F.}\ \bibnamefont
  {Degtyareva}},\ }\href {\doibase 10.1080/0895795032000102441} {\bibfield
  {journal} {\bibinfo  {journal} {High Press. Res.}\ }\textbf {\bibinfo
  {volume} {23}},\ \bibinfo {pages} {253} (\bibinfo {year} {2003})}\BibitemShut
  {NoStop}%
\bibitem [{\citenamefont {Ackland}\ and\ \citenamefont
  {Macleod}(2004)}]{Ackland2004}%
  \BibitemOpen
  \bibfield  {author} {\bibinfo {author} {\bibfnamefont {G.~J.}\ \bibnamefont
  {Ackland}}\ and\ \bibinfo {author} {\bibfnamefont {I.~R.}\ \bibnamefont
  {Macleod}},\ }\href {\doibase 10.1088/1367-2630/6/1/138} {\bibfield
  {journal} {\bibinfo  {journal} {New J. Phys.}\ }\textbf {\bibinfo {volume}
  {6}},\ \bibinfo {pages} {138} (\bibinfo {year} {2004})}\BibitemShut {NoStop}%
\bibitem [{\citenamefont {Katzke}\ and\ \citenamefont
  {Tol\'{e}dano}(2005)}]{Katzke2005}%
  \BibitemOpen
  \bibfield  {author} {\bibinfo {author} {\bibfnamefont {H.}~\bibnamefont
  {Katzke}}\ and\ \bibinfo {author} {\bibfnamefont {P.}~\bibnamefont
  {Tol\'{e}dano}},\ }\href {\doibase 10.1103/PhysRevB.71.184101} {\bibfield
  {journal} {\bibinfo  {journal} {Phys. Rev. B}\ }\textbf {\bibinfo {volume}
  {71}},\ \bibinfo {pages} {184101} (\bibinfo {year} {2005})}\BibitemShut
  {NoStop}%
\bibitem [{\citenamefont {McMahon}\ \emph {et~al.}(2006)\citenamefont
  {McMahon}, \citenamefont {Nelmes}, \citenamefont {Schwarz},\ and\
  \citenamefont {Syassen}}]{McMahon2006}%
  \BibitemOpen
  \bibfield  {author} {\bibinfo {author} {\bibfnamefont {M.}~\bibnamefont
  {McMahon}}, \bibinfo {author} {\bibfnamefont {R.}~\bibnamefont {Nelmes}},
  \bibinfo {author} {\bibfnamefont {U.}~\bibnamefont {Schwarz}}, \ and\
  \bibinfo {author} {\bibfnamefont {K.}~\bibnamefont {Syassen}},\ }\href
  {\doibase 10.1103/PhysRevB.74.140102} {\bibfield  {journal} {\bibinfo
  {journal} {Phys. Rev. B}\ }\textbf {\bibinfo {volume} {74}},\ \bibinfo
  {pages} {140102} (\bibinfo {year} {2006})}\BibitemShut {NoStop}%
\bibitem [{\citenamefont {Falconi}\ \emph {et~al.}(2006)\citenamefont
  {Falconi}, \citenamefont {McMahon}, \citenamefont {Lundegaard}, \citenamefont
  {Hejny}, \citenamefont {Nelmes},\ and\ \citenamefont
  {Hanfland}}]{Falconi2006}%
  \BibitemOpen
  \bibfield  {author} {\bibinfo {author} {\bibfnamefont {S.}~\bibnamefont
  {Falconi}}, \bibinfo {author} {\bibfnamefont {M.}~\bibnamefont {McMahon}},
  \bibinfo {author} {\bibfnamefont {L.}~\bibnamefont {Lundegaard}}, \bibinfo
  {author} {\bibfnamefont {C.}~\bibnamefont {Hejny}}, \bibinfo {author}
  {\bibfnamefont {R.}~\bibnamefont {Nelmes}}, \ and\ \bibinfo {author}
  {\bibfnamefont {M.}~\bibnamefont {Hanfland}},\ }\href {\doibase
  10.1103/PhysRevB.73.214102} {\bibfield  {journal} {\bibinfo  {journal} {Phys.
  Rev. B}\ }\textbf {\bibinfo {volume} {73}},\ \bibinfo {pages} {214102}
  (\bibinfo {year} {2006})}\BibitemShut {NoStop}%
\bibitem [{\citenamefont {Marqu\'{e}s}\ \emph {et~al.}(2009)\citenamefont
  {Marqu\'{e}s}, \citenamefont {Ackland}, \citenamefont {Lundegaard},
  \citenamefont {Stinton}, \citenamefont {Nelmes}, \citenamefont {McMahon},\
  and\ \citenamefont {Contreras-Garc\'{\i}a}}]{Marques2009}%
  \BibitemOpen
  \bibfield  {author} {\bibinfo {author} {\bibfnamefont {M.}~\bibnamefont
  {Marqu\'{e}s}}, \bibinfo {author} {\bibfnamefont {G.}~\bibnamefont
  {Ackland}}, \bibinfo {author} {\bibfnamefont {L.}~\bibnamefont {Lundegaard}},
  \bibinfo {author} {\bibfnamefont {G.}~\bibnamefont {Stinton}}, \bibinfo
  {author} {\bibfnamefont {R.}~\bibnamefont {Nelmes}}, \bibinfo {author}
  {\bibfnamefont {M.}~\bibnamefont {McMahon}}, \ and\ \bibinfo {author}
  {\bibfnamefont {J.}~\bibnamefont {Contreras-Garc\'{\i}a}},\ }\href {\doibase
  10.1103/PhysRevLett.103.115501} {\bibfield  {journal} {\bibinfo  {journal}
  {Phys. Rev. Lett.}\ }\textbf {\bibinfo {volume} {103}},\ \bibinfo {pages}
  {115501} (\bibinfo {year} {2009})}\BibitemShut {NoStop}%
\bibitem [{\citenamefont {Lundegaard}\ \emph {et~al.}(2009)\citenamefont
  {Lundegaard}, \citenamefont {Marqu\'{e}s}, \citenamefont {Stinton},
  \citenamefont {Ackland}, \citenamefont {Nelmes},\ and\ \citenamefont
  {McMahon}}]{Lundegaard2009}%
  \BibitemOpen
  \bibfield  {author} {\bibinfo {author} {\bibfnamefont {L.}~\bibnamefont
  {Lundegaard}}, \bibinfo {author} {\bibfnamefont {M.}~\bibnamefont
  {Marqu\'{e}s}}, \bibinfo {author} {\bibfnamefont {G.}~\bibnamefont
  {Stinton}}, \bibinfo {author} {\bibfnamefont {G.}~\bibnamefont {Ackland}},
  \bibinfo {author} {\bibfnamefont {R.}~\bibnamefont {Nelmes}}, \ and\ \bibinfo
  {author} {\bibfnamefont {M.}~\bibnamefont {McMahon}},\ }\href {\doibase
  10.1103/PhysRevB.80.020101} {\bibfield  {journal} {\bibinfo  {journal} {Phys.
  Rev. B}\ }\textbf {\bibinfo {volume} {80}},\ \bibinfo {pages} {020101}
  (\bibinfo {year} {2009})}\BibitemShut {NoStop}%
\bibitem [{\citenamefont {Degtyareva}\ and\ \citenamefont
  {Degtyareva}(2009)}]{Degtyareva2009}%
  \BibitemOpen
  \bibfield  {author} {\bibinfo {author} {\bibfnamefont {V.~F.}\ \bibnamefont
  {Degtyareva}}\ and\ \bibinfo {author} {\bibfnamefont {O.}~\bibnamefont
  {Degtyareva}},\ }\href {\doibase 10.1088/1367-2630/11/6/063037} {\bibfield
  {journal} {\bibinfo  {journal} {New J. Phys.}\ }\textbf {\bibinfo {volume}
  {11}},\ \bibinfo {pages} {063037} (\bibinfo {year} {2009})}\BibitemShut
  {NoStop}%
\bibitem [{\citenamefont {Pickard}\ and\ \citenamefont
  {Needs}(2011)}]{Pickard2011}%
  \BibitemOpen
  \bibfield  {author} {\bibinfo {author} {\bibfnamefont {C.~J.}\ \bibnamefont
  {Pickard}}\ and\ \bibinfo {author} {\bibfnamefont {R.~J.}\ \bibnamefont
  {Needs}},\ }\href {\doibase 10.1103/PhysRevLett.107.087201} {\bibfield
  {journal} {\bibinfo  {journal} {Phys. Rev. Lett.}\ }\textbf {\bibinfo
  {volume} {107}},\ \bibinfo {pages} {087201} (\bibinfo {year}
  {2011})}\BibitemShut {NoStop}%
\bibitem [{\citenamefont {Guillaume}\ \emph {et~al.}(2011)\citenamefont
  {Guillaume}, \citenamefont {Gregoryanz}, \citenamefont {Degtyareva},
  \citenamefont {McMahon}, \citenamefont {Hanfland}, \citenamefont {Evans},
  \citenamefont {Guthrie}, \citenamefont {Sinogeikin},\ and\ \citenamefont
  {Mao}}]{Guillaume2011}%
  \BibitemOpen
  \bibfield  {author} {\bibinfo {author} {\bibfnamefont {C.~L.}\ \bibnamefont
  {Guillaume}}, \bibinfo {author} {\bibfnamefont {E.}~\bibnamefont
  {Gregoryanz}}, \bibinfo {author} {\bibfnamefont {O.}~\bibnamefont
  {Degtyareva}}, \bibinfo {author} {\bibfnamefont {M.~I.}\ \bibnamefont
  {McMahon}}, \bibinfo {author} {\bibfnamefont {M.}~\bibnamefont {Hanfland}},
  \bibinfo {author} {\bibfnamefont {S.}~\bibnamefont {Evans}}, \bibinfo
  {author} {\bibfnamefont {M.}~\bibnamefont {Guthrie}}, \bibinfo {author}
  {\bibfnamefont {S.~V.}\ \bibnamefont {Sinogeikin}}, \ and\ \bibinfo {author}
  {\bibfnamefont {H.-K.}\ \bibnamefont {Mao}},\ }\href {\doibase
  10.1038/nphys1864} {\bibfield  {journal} {\bibinfo  {journal} {Nat. Phys.}\
  }\textbf {\bibinfo {volume} {7}},\ \bibinfo {pages} {211} (\bibinfo {year}
  {2011})}\BibitemShut {NoStop}%
\bibitem [{\citenamefont {Jones}(1934)}]{Jones1934}%
  \BibitemOpen
  \bibfield  {author} {\bibinfo {author} {\bibfnamefont {H.}~\bibnamefont
  {Jones}},\ }\href {\doibase 10.1098/rspa.1934.0224} {\bibfield  {journal}
  {\bibinfo  {journal} {Proc. R. Soc. A Math. Phys. Eng. Sci.}\ }\textbf
  {\bibinfo {volume} {147}},\ \bibinfo {pages} {396} (\bibinfo {year}
  {1934})}\BibitemShut {NoStop}%
\bibitem [{\citenamefont {Mott}\ and\ \citenamefont {Jones}(1936)}]{Mott1936}%
  \BibitemOpen
  \bibfield  {author} {\bibinfo {author} {\bibfnamefont {N.~F.}\ \bibnamefont
  {Mott}}\ and\ \bibinfo {author} {\bibfnamefont {H.}~\bibnamefont {Jones}},\
  }\href@noop {} {\emph {\bibinfo {title} {{The Theory of the Properties of
  Metals and Alloys}}}}\ (\bibinfo  {publisher} {Oxford Univeristy Press},\
  \bibinfo {address} {London},\ \bibinfo {year} {1936})\BibitemShut {NoStop}%
\bibitem [{\citenamefont {Neaton}\ and\ \citenamefont
  {Ashcroft}(1999)}]{Neaton1999}%
  \BibitemOpen
  \bibfield  {author} {\bibinfo {author} {\bibfnamefont {J.~B.}\ \bibnamefont
  {Neaton}}\ and\ \bibinfo {author} {\bibfnamefont {N.~W.}\ \bibnamefont
  {Ashcroft}},\ }\href {\doibase 10.1038/22067} {\bibfield  {journal} {\bibinfo
   {journal} {Nature}\ }\textbf {\bibinfo {volume} {400}},\ \bibinfo {pages}
  {141} (\bibinfo {year} {1999})}\BibitemShut {NoStop}%
\bibitem [{\citenamefont {Neaton}\ and\ \citenamefont
  {Ashcroft}(2001)}]{Neaton2001}%
  \BibitemOpen
  \bibfield  {author} {\bibinfo {author} {\bibfnamefont {J.~B.}\ \bibnamefont
  {Neaton}}\ and\ \bibinfo {author} {\bibfnamefont {N.~W.}\ \bibnamefont
  {Ashcroft}},\ }\href {\doibase 10.1103/PhysRevLett.86.2830} {\bibfield
  {journal} {\bibinfo  {journal} {Phys. Rev. Lett.}\ }\textbf {\bibinfo
  {volume} {86}},\ \bibinfo {pages} {2830} (\bibinfo {year}
  {2001})}\BibitemShut {NoStop}%
\bibitem [{\citenamefont {Rousseau}\ and\ \citenamefont
  {Ashcroft}(2008)}]{Rousseau2008}%
  \BibitemOpen
  \bibfield  {author} {\bibinfo {author} {\bibfnamefont {B.}~\bibnamefont
  {Rousseau}}\ and\ \bibinfo {author} {\bibfnamefont {N.~W.}\ \bibnamefont
  {Ashcroft}},\ }\href {\doibase 10.1103/PhysRevLett.101.046407} {\bibfield
  {journal} {\bibinfo  {journal} {Phys. Rev. Lett.}\ }\textbf {\bibinfo
  {volume} {101}},\ \bibinfo {pages} {046407} (\bibinfo {year}
  {2008})}\BibitemShut {NoStop}%
\bibitem [{\citenamefont {Xie}\ \emph {et~al.}(2007)\citenamefont {Xie},
  \citenamefont {Tse}, \citenamefont {Cui}, \citenamefont {Oganov},
  \citenamefont {He}, \citenamefont {Ma},\ and\ \citenamefont {Zou}}]{Xie2007}%
  \BibitemOpen
  \bibfield  {author} {\bibinfo {author} {\bibfnamefont {Y.}~\bibnamefont
  {Xie}}, \bibinfo {author} {\bibfnamefont {J.}~\bibnamefont {Tse}}, \bibinfo
  {author} {\bibfnamefont {T.}~\bibnamefont {Cui}}, \bibinfo {author}
  {\bibfnamefont {A.}~\bibnamefont {Oganov}}, \bibinfo {author} {\bibfnamefont
  {Z.}~\bibnamefont {He}}, \bibinfo {author} {\bibfnamefont {Y.}~\bibnamefont
  {Ma}}, \ and\ \bibinfo {author} {\bibfnamefont {G.}~\bibnamefont {Zou}},\
  }\href {\doibase 10.1103/PhysRevB.75.064102} {\bibfield  {journal} {\bibinfo
  {journal} {Phys. Rev. B}\ }\textbf {\bibinfo {volume} {75}},\ \bibinfo
  {pages} {064102} (\bibinfo {year} {2007})}\BibitemShut {NoStop}%
\bibitem [{\citenamefont {S\"{o}derlind}\ \emph {et~al.}(1995)\citenamefont
  {S\"{o}derlind}, \citenamefont {Eriksson}, \citenamefont {Johansson},
  \citenamefont {Wills},\ and\ \citenamefont {Boring}}]{Soderlind1995}%
  \BibitemOpen
  \bibfield  {author} {\bibinfo {author} {\bibfnamefont {P.}~\bibnamefont
  {S\"{o}derlind}}, \bibinfo {author} {\bibfnamefont {O.}~\bibnamefont
  {Eriksson}}, \bibinfo {author} {\bibfnamefont {B.}~\bibnamefont {Johansson}},
  \bibinfo {author} {\bibfnamefont {J.~M.}\ \bibnamefont {Wills}}, \ and\
  \bibinfo {author} {\bibfnamefont {A.~M.}\ \bibnamefont {Boring}},\ }\href
  {\doibase 10.1038/374524a0} {\bibfield  {journal} {\bibinfo  {journal}
  {Nature}\ }\textbf {\bibinfo {volume} {374}},\ \bibinfo {pages} {524}
  (\bibinfo {year} {1995})}\BibitemShut {NoStop}%
\bibitem [{\citenamefont {Moore}\ and\ \citenamefont {van~der
  Laan}(2009)}]{Moore2009}%
  \BibitemOpen
  \bibfield  {author} {\bibinfo {author} {\bibfnamefont {K.}~\bibnamefont
  {Moore}}\ and\ \bibinfo {author} {\bibfnamefont {G.}~\bibnamefont {van~der
  Laan}},\ }\href {\doibase 10.1103/RevModPhys.81.235} {\bibfield  {journal}
  {\bibinfo  {journal} {Rev. Mod. Phys.}\ }\textbf {\bibinfo {volume} {81}},\
  \bibinfo {pages} {235} (\bibinfo {year} {2009})}\BibitemShut {NoStop}%
\bibitem [{\citenamefont {Boehler}\ and\ \citenamefont
  {Zha}(1986)}]{Boehler1986}%
  \BibitemOpen
  \bibfield  {author} {\bibinfo {author} {\bibfnamefont {R.}~\bibnamefont
  {Boehler}}\ and\ \bibinfo {author} {\bibfnamefont {C.-S.}\ \bibnamefont
  {Zha}},\ }\href {\doibase 10.1016/0378-4363(86)90565-6} {\bibfield  {journal}
  {\bibinfo  {journal} {Phys. B+C}\ }\textbf {\bibinfo {volume} {139-140}},\
  \bibinfo {pages} {233} (\bibinfo {year} {1986})}\BibitemShut {NoStop}%
\bibitem [{\citenamefont {Gregoryanz}\ \emph {et~al.}(2005)\citenamefont
  {Gregoryanz}, \citenamefont {Degtyareva}, \citenamefont {Somayazulu},
  \citenamefont {Hemley},\ and\ \citenamefont {Mao}}]{Gregoryanz2005}%
  \BibitemOpen
  \bibfield  {author} {\bibinfo {author} {\bibfnamefont {E.}~\bibnamefont
  {Gregoryanz}}, \bibinfo {author} {\bibfnamefont {O.}~\bibnamefont
  {Degtyareva}}, \bibinfo {author} {\bibfnamefont {M.}~\bibnamefont
  {Somayazulu}}, \bibinfo {author} {\bibfnamefont {R.}~\bibnamefont {Hemley}},
  \ and\ \bibinfo {author} {\bibfnamefont {H.-k.}\ \bibnamefont {Mao}},\ }\href
  {\doibase 10.1103/PhysRevLett.94.185502} {\bibfield  {journal} {\bibinfo
  {journal} {Phys. Rev. Lett.}\ }\textbf {\bibinfo {volume} {94}},\ \bibinfo
  {pages} {185502} (\bibinfo {year} {2005})}\BibitemShut {NoStop}%
\bibitem [{\citenamefont {Schaeffer}\ \emph {et~al.}(2012)\citenamefont
  {Schaeffer}, \citenamefont {Talmadge}, \citenamefont {Temple},\ and\
  \citenamefont {Deemyad}}]{Schaeffer2012}%
  \BibitemOpen
  \bibfield  {author} {\bibinfo {author} {\bibfnamefont {A.~M.~J.}\
  \bibnamefont {Schaeffer}}, \bibinfo {author} {\bibfnamefont {W.~B.}\
  \bibnamefont {Talmadge}}, \bibinfo {author} {\bibfnamefont {S.~R.}\
  \bibnamefont {Temple}}, \ and\ \bibinfo {author} {\bibfnamefont
  {S.}~\bibnamefont {Deemyad}},\ }\href {\doibase
  10.1103/PhysRevLett.109.185702} {\bibfield  {journal} {\bibinfo  {journal}
  {Phys. Rev. Lett.}\ }\textbf {\bibinfo {volume} {109}},\ \bibinfo {pages}
  {185702} (\bibinfo {year} {2012})}\BibitemShut {NoStop}%
\bibitem [{\citenamefont {Loa}\ \emph {et~al.}(2011)\citenamefont {Loa},
  \citenamefont {Syassen}, \citenamefont {Monaco}, \citenamefont {Vank\'{o}},
  \citenamefont {Krisch},\ and\ \citenamefont {Hanfland}}]{Loa2011}%
  \BibitemOpen
  \bibfield  {author} {\bibinfo {author} {\bibfnamefont {I.}~\bibnamefont
  {Loa}}, \bibinfo {author} {\bibfnamefont {K.}~\bibnamefont {Syassen}},
  \bibinfo {author} {\bibfnamefont {G.}~\bibnamefont {Monaco}}, \bibinfo
  {author} {\bibfnamefont {G.}~\bibnamefont {Vank\'{o}}}, \bibinfo {author}
  {\bibfnamefont {M.}~\bibnamefont {Krisch}}, \ and\ \bibinfo {author}
  {\bibfnamefont {M.}~\bibnamefont {Hanfland}},\ }\href {\doibase
  10.1103/PhysRevLett.107.086402} {\bibfield  {journal} {\bibinfo  {journal}
  {Phys. Rev. Lett.}\ }\textbf {\bibinfo {volume} {107}},\ \bibinfo {pages}
  {086402} (\bibinfo {year} {2011})}\BibitemShut {NoStop}%
\bibitem [{\citenamefont {Xie}\ \emph {et~al.}(2000)\citenamefont {Xie},
  \citenamefont {Chen}, \citenamefont {Tse}, \citenamefont {Klug},
  \citenamefont {Li}, \citenamefont {Uehara},\ and\ \citenamefont
  {Wang}}]{Xie2000}%
  \BibitemOpen
  \bibfield  {author} {\bibinfo {author} {\bibfnamefont {J.}~\bibnamefont
  {Xie}}, \bibinfo {author} {\bibfnamefont {S.}~\bibnamefont {Chen}}, \bibinfo
  {author} {\bibfnamefont {J.}~\bibnamefont {Tse}}, \bibinfo {author}
  {\bibfnamefont {D.}~\bibnamefont {Klug}}, \bibinfo {author} {\bibfnamefont
  {Z.}~\bibnamefont {Li}}, \bibinfo {author} {\bibfnamefont {K.}~\bibnamefont
  {Uehara}}, \ and\ \bibinfo {author} {\bibfnamefont {L.}~\bibnamefont
  {Wang}},\ }\href {\doibase 10.1103/PhysRevB.62.3624} {\bibfield  {journal}
  {\bibinfo  {journal} {Phys. Rev. B}\ }\textbf {\bibinfo {volume} {62}},\
  \bibinfo {pages} {3624} (\bibinfo {year} {2000})}\BibitemShut {NoStop}%
\bibitem [{\citenamefont {Rodriguez-Prieto}\ \emph {et~al.}(2006)\citenamefont
  {Rodriguez-Prieto}, \citenamefont {Bergara}, \citenamefont {Silkin},\ and\
  \citenamefont {Echenique}}]{Rodriguez-Prieto2006}%
  \BibitemOpen
  \bibfield  {author} {\bibinfo {author} {\bibfnamefont {A.}~\bibnamefont
  {Rodriguez-Prieto}}, \bibinfo {author} {\bibfnamefont {A.}~\bibnamefont
  {Bergara}}, \bibinfo {author} {\bibfnamefont {V.}~\bibnamefont {Silkin}}, \
  and\ \bibinfo {author} {\bibfnamefont {P.}~\bibnamefont {Echenique}},\ }\href
  {\doibase 10.1103/PhysRevB.74.172104} {\bibfield  {journal} {\bibinfo
  {journal} {Phys. Rev. B}\ }\textbf {\bibinfo {volume} {74}},\ \bibinfo
  {pages} {172104} (\bibinfo {year} {2006})}\BibitemShut {NoStop}%
\bibitem [{\citenamefont {Xie}\ \emph {et~al.}(2008)\citenamefont {Xie},
  \citenamefont {Ma}, \citenamefont {Cui}, \citenamefont {Li}, \citenamefont
  {Qiu},\ and\ \citenamefont {Zou}}]{Xie2008}%
  \BibitemOpen
  \bibfield  {author} {\bibinfo {author} {\bibfnamefont {Y.}~\bibnamefont
  {Xie}}, \bibinfo {author} {\bibfnamefont {Y.~M.}\ \bibnamefont {Ma}},
  \bibinfo {author} {\bibfnamefont {T.}~\bibnamefont {Cui}}, \bibinfo {author}
  {\bibfnamefont {Y.}~\bibnamefont {Li}}, \bibinfo {author} {\bibfnamefont
  {J.}~\bibnamefont {Qiu}}, \ and\ \bibinfo {author} {\bibfnamefont {G.~T.}\
  \bibnamefont {Zou}},\ }\href {\doibase 10.1088/1367-2630/10/6/063022}
  {\bibfield  {journal} {\bibinfo  {journal} {New J. Phys.}\ }\textbf {\bibinfo
  {volume} {10}},\ \bibinfo {pages} {063022} (\bibinfo {year}
  {2008})}\BibitemShut {NoStop}%
\bibitem [{\citenamefont {Hume-Rothery}(1926)}]{Hume-Rothery1926}%
  \BibitemOpen
  \bibfield  {author} {\bibinfo {author} {\bibfnamefont {W.}~\bibnamefont
  {Hume-Rothery}},\ }\href@noop {} {\bibfield  {journal} {\bibinfo  {journal}
  {J. Inst. Met.}\ }\textbf {\bibinfo {volume} {35}},\ \bibinfo {pages} {319}
  (\bibinfo {year} {1926})}\BibitemShut {NoStop}%
\bibitem [{\citenamefont {Degtyareva}(2006)}]{Degtyareva2006}%
  \BibitemOpen
  \bibfield  {author} {\bibinfo {author} {\bibfnamefont {V.}~\bibnamefont
  {Degtyareva}},\ }\href {\doibase 10.1070/PU2006v049n04ABEH005948} {\bibfield
  {journal} {\bibinfo  {journal} {Physics-Uspekhi}\ }\textbf {\bibinfo {volume}
  {49}},\ \bibinfo {pages} {369} (\bibinfo {year} {2006})}\BibitemShut
  {NoStop}%
\bibitem [{\citenamefont {Degtyareva}(2014)}]{Degtyareva2014}%
  \BibitemOpen
  \bibfield  {author} {\bibinfo {author} {\bibfnamefont {V.}~\bibnamefont
  {Degtyareva}},\ }\href {\doibase 10.1016/j.solidstatesciences.2014.07.008}
  {\bibfield  {journal} {\bibinfo  {journal} {Solid State Sci.}\ }\textbf
  {\bibinfo {volume} {36}},\ \bibinfo {pages} {62} (\bibinfo {year}
  {2014})}\BibitemShut {NoStop}%
\bibitem [{\citenamefont {Stoner}(1939)}]{Stoner1939}%
  \BibitemOpen
  \bibfield  {author} {\bibinfo {author} {\bibfnamefont {E.~C.}\ \bibnamefont
  {Stoner}},\ }\href {\doibase 10.1098/rspa.1939.0003} {\bibfield  {journal}
  {\bibinfo  {journal} {Proc. R. Soc. A Math. Phys. Eng. Sci.}\ }\textbf
  {\bibinfo {volume} {169}},\ \bibinfo {pages} {339} (\bibinfo {year}
  {1939})}\BibitemShut {NoStop}%
\bibitem [{\citenamefont {Abd-Elmeguid}\ \emph {et~al.}(1994)\citenamefont
  {Abd-Elmeguid}, \citenamefont {Pattyn},\ and\ \citenamefont
  {Bukshpan}}]{Abd-Elmeguid1994}%
  \BibitemOpen
  \bibfield  {author} {\bibinfo {author} {\bibfnamefont {M.}~\bibnamefont
  {Abd-Elmeguid}}, \bibinfo {author} {\bibfnamefont {H.}~\bibnamefont
  {Pattyn}}, \ and\ \bibinfo {author} {\bibfnamefont {S.}~\bibnamefont
  {Bukshpan}},\ }\href {\doibase 10.1103/PhysRevLett.72.502} {\bibfield
  {journal} {\bibinfo  {journal} {Phys. Rev. Lett.}\ }\textbf {\bibinfo
  {volume} {72}},\ \bibinfo {pages} {502} (\bibinfo {year} {1994})}\BibitemShut
  {NoStop}%
\bibitem [{\citenamefont {Tomita}\ \emph {et~al.}(2005)\citenamefont {Tomita},
  \citenamefont {Deemyad}, \citenamefont {Hamlin}, \citenamefont {Schilling},
  \citenamefont {Tissen}, \citenamefont {Veal}, \citenamefont {Chen},\ and\
  \citenamefont {Claus}}]{Tomita2005}%
  \BibitemOpen
  \bibfield  {author} {\bibinfo {author} {\bibfnamefont {T.}~\bibnamefont
  {Tomita}}, \bibinfo {author} {\bibfnamefont {S.}~\bibnamefont {Deemyad}},
  \bibinfo {author} {\bibfnamefont {J.~J.}\ \bibnamefont {Hamlin}}, \bibinfo
  {author} {\bibfnamefont {J.~S.}\ \bibnamefont {Schilling}}, \bibinfo {author}
  {\bibfnamefont {V.~G.}\ \bibnamefont {Tissen}}, \bibinfo {author}
  {\bibfnamefont {B.~W.}\ \bibnamefont {Veal}}, \bibinfo {author}
  {\bibfnamefont {L.}~\bibnamefont {Chen}}, \ and\ \bibinfo {author}
  {\bibfnamefont {H.}~\bibnamefont {Claus}},\ }\href {\doibase
  10.1088/0953-8984/17/11/024} {\bibfield  {journal} {\bibinfo  {journal} {J.
  Phys. Condens. Matter}\ }\textbf {\bibinfo {volume} {17}},\ \bibinfo {pages}
  {S921} (\bibinfo {year} {2005})}\BibitemShut {NoStop}%
\bibitem [{\citenamefont {Schilling}(2006)}]{Schilling2006}%
  \BibitemOpen
  \bibfield  {author} {\bibinfo {author} {\bibfnamefont {J.~S.}\ \bibnamefont
  {Schilling}},\ }\href {\doibase 10.1080/08957950600864401} {\bibfield
  {journal} {\bibinfo  {journal} {High Press. Res.}\ }\textbf {\bibinfo
  {volume} {26}},\ \bibinfo {pages} {145} (\bibinfo {year} {2006})}\BibitemShut
  {NoStop}%
\bibitem [{\citenamefont {Louie}\ and\ \citenamefont
  {Cohen}(1974)}]{Louie1974}%
  \BibitemOpen
  \bibfield  {author} {\bibinfo {author} {\bibfnamefont {S.}~\bibnamefont
  {Louie}}\ and\ \bibinfo {author} {\bibfnamefont {M.}~\bibnamefont {Cohen}},\
  }\href {\doibase 10.1103/PhysRevB.10.3237} {\bibfield  {journal} {\bibinfo
  {journal} {Phys. Rev. B}\ }\textbf {\bibinfo {volume} {10}},\ \bibinfo
  {pages} {3237} (\bibinfo {year} {1974})}\BibitemShut {NoStop}%
\bibitem [{\citenamefont {McMahan}(1978)}]{McMahan1978}%
  \BibitemOpen
  \bibfield  {author} {\bibinfo {author} {\bibfnamefont {A.}~\bibnamefont
  {McMahan}},\ }\href {\doibase 10.1103/PhysRevB.17.1521} {\bibfield  {journal}
  {\bibinfo  {journal} {Phys. Rev. B}\ }\textbf {\bibinfo {volume} {17}},\
  \bibinfo {pages} {1521} (\bibinfo {year} {1978})}\BibitemShut {NoStop}%
\bibitem [{\citenamefont {McMahan}(1984)}]{McMahan1984}%
  \BibitemOpen
  \bibfield  {author} {\bibinfo {author} {\bibfnamefont {A.}~\bibnamefont
  {McMahan}},\ }\href {\doibase 10.1103/PhysRevB.29.5982} {\bibfield  {journal}
  {\bibinfo  {journal} {Phys. Rev. B}\ }\textbf {\bibinfo {volume} {29}},\
  \bibinfo {pages} {5982} (\bibinfo {year} {1984})}\BibitemShut {NoStop}%
\bibitem [{\citenamefont {Young}\ and\ \citenamefont {Ross}(1984)}]{Young1984}%
  \BibitemOpen
  \bibfield  {author} {\bibinfo {author} {\bibfnamefont {D.}~\bibnamefont
  {Young}}\ and\ \bibinfo {author} {\bibfnamefont {M.}~\bibnamefont {Ross}},\
  }\href {\doibase 10.1103/PhysRevB.29.682} {\bibfield  {journal} {\bibinfo
  {journal} {Phys. Rev. B}\ }\textbf {\bibinfo {volume} {29}},\ \bibinfo
  {pages} {682} (\bibinfo {year} {1984})}\BibitemShut {NoStop}%
\bibitem [{\citenamefont {Skriver}(1985)}]{Skriver1985}%
  \BibitemOpen
  \bibfield  {author} {\bibinfo {author} {\bibfnamefont {H.~L.}\ \bibnamefont
  {Skriver}},\ }\href {\doibase 10.1103/PhysRevB.31.1909} {\bibfield  {journal}
  {\bibinfo  {journal} {Phys. Rev. B}\ }\textbf {\bibinfo {volume} {31}},\
  \bibinfo {pages} {1909} (\bibinfo {year} {1985})}\BibitemShut {NoStop}%
\bibitem [{\citenamefont {Ahuja}\ \emph {et~al.}(2000)\citenamefont {Ahuja},
  \citenamefont {Eriksson},\ and\ \citenamefont {Johansson}}]{Ahuja2000}%
  \BibitemOpen
  \bibfield  {author} {\bibinfo {author} {\bibfnamefont {R.}~\bibnamefont
  {Ahuja}}, \bibinfo {author} {\bibfnamefont {O.}~\bibnamefont {Eriksson}}, \
  and\ \bibinfo {author} {\bibfnamefont {B.}~\bibnamefont {Johansson}},\ }\href
  {\doibase 10.1103/PhysRevB.63.014102} {\bibfield  {journal} {\bibinfo
  {journal} {Phys. Rev. B}\ }\textbf {\bibinfo {volume} {63}},\ \bibinfo
  {pages} {014102} (\bibinfo {year} {2000})}\BibitemShut {NoStop}%
\bibitem [{\citenamefont {Schwarz}\ \emph {et~al.}(2000)\citenamefont
  {Schwarz}, \citenamefont {Jepsen},\ and\ \citenamefont
  {Syassen}}]{Schwarz2000}%
  \BibitemOpen
  \bibfield  {author} {\bibinfo {author} {\bibfnamefont {U.}~\bibnamefont
  {Schwarz}}, \bibinfo {author} {\bibfnamefont {O.}~\bibnamefont {Jepsen}}, \
  and\ \bibinfo {author} {\bibfnamefont {K.}~\bibnamefont {Syassen}},\ }\href
  {\doibase 10.1016/S0038-1098(99)00527-X} {\bibfield  {journal} {\bibinfo
  {journal} {Solid State Commun.}\ }\textbf {\bibinfo {volume} {113}},\
  \bibinfo {pages} {643} (\bibinfo {year} {2000})}\BibitemShut {NoStop}%
\bibitem [{\citenamefont {Shi}\ and\ \citenamefont
  {Papaconstantopoulos}(2006)}]{Shi2006}%
  \BibitemOpen
  \bibfield  {author} {\bibinfo {author} {\bibfnamefont {L.}~\bibnamefont
  {Shi}}\ and\ \bibinfo {author} {\bibfnamefont {D.}~\bibnamefont
  {Papaconstantopoulos}},\ }\href {\doibase 10.1103/PhysRevB.73.184516}
  {\bibfield  {journal} {\bibinfo  {journal} {Phys. Rev. B}\ }\textbf {\bibinfo
  {volume} {73}},\ \bibinfo {pages} {184516} (\bibinfo {year}
  {2006})}\BibitemShut {NoStop}%
\bibitem [{\citenamefont {Profeta}\ \emph {et~al.}(2006)\citenamefont
  {Profeta}, \citenamefont {Franchini}, \citenamefont {Lathiotakis},
  \citenamefont {Floris}, \citenamefont {Sanna}, \citenamefont {Marques},
  \citenamefont {L\"{u}ders}, \citenamefont {Massidda}, \citenamefont {Gross},\
  and\ \citenamefont {Continenza}}]{Profeta2006}%
  \BibitemOpen
  \bibfield  {author} {\bibinfo {author} {\bibfnamefont {G.}~\bibnamefont
  {Profeta}}, \bibinfo {author} {\bibfnamefont {C.}~\bibnamefont {Franchini}},
  \bibinfo {author} {\bibfnamefont {N.}~\bibnamefont {Lathiotakis}}, \bibinfo
  {author} {\bibfnamefont {a.}~\bibnamefont {Floris}}, \bibinfo {author}
  {\bibfnamefont {a.}~\bibnamefont {Sanna}}, \bibinfo {author} {\bibfnamefont
  {M.}~\bibnamefont {Marques}}, \bibinfo {author} {\bibfnamefont
  {M.}~\bibnamefont {L\"{u}ders}}, \bibinfo {author} {\bibfnamefont
  {S.}~\bibnamefont {Massidda}}, \bibinfo {author} {\bibfnamefont
  {E.}~\bibnamefont {Gross}}, \ and\ \bibinfo {author} {\bibfnamefont
  {a.}~\bibnamefont {Continenza}},\ }\href {\doibase
  10.1103/PhysRevLett.96.047003} {\bibfield  {journal} {\bibinfo  {journal}
  {Phys. Rev. Lett.}\ }\textbf {\bibinfo {volume} {96}},\ \bibinfo {pages}
  {047003} (\bibinfo {year} {2006})}\BibitemShut {NoStop}%
\bibitem [{\citenamefont {Perez-Mato}\ \emph {et~al.}(2007)\citenamefont
  {Perez-Mato}, \citenamefont {Elcoro}, \citenamefont {Pet{\v r}\'{\i}\v{c}ek},
  \citenamefont {Katzke},\ and\ \citenamefont {Blaha}}]{Perez-Mato2007}%
  \BibitemOpen
  \bibfield  {author} {\bibinfo {author} {\bibfnamefont {J.}~\bibnamefont
  {Perez-Mato}}, \bibinfo {author} {\bibfnamefont {L.}~\bibnamefont {Elcoro}},
  \bibinfo {author} {\bibfnamefont {V.}~\bibnamefont {Pet{\v r}\'{\i}\v{c}ek}},
  \bibinfo {author} {\bibfnamefont {H.}~\bibnamefont {Katzke}}, \ and\ \bibinfo
  {author} {\bibfnamefont {P.}~\bibnamefont {Blaha}},\ }\href {\doibase
  10.1103/PhysRevLett.99.025502} {\bibfield  {journal} {\bibinfo  {journal}
  {Phys. Rev. Lett.}\ }\textbf {\bibinfo {volume} {99}},\ \bibinfo {pages}
  {025502} (\bibinfo {year} {2007})}\BibitemShut {NoStop}%
\bibitem [{\citenamefont {Ma}\ \emph {et~al.}(2008)\citenamefont {Ma},
  \citenamefont {Oganov},\ and\ \citenamefont {Xie}}]{Ma2008}%
  \BibitemOpen
  \bibfield  {author} {\bibinfo {author} {\bibfnamefont {Y.}~\bibnamefont
  {Ma}}, \bibinfo {author} {\bibfnamefont {A.}~\bibnamefont {Oganov}}, \ and\
  \bibinfo {author} {\bibfnamefont {Y.}~\bibnamefont {Xie}},\ }\href {\doibase
  10.1103/PhysRevB.78.014102} {\bibfield  {journal} {\bibinfo  {journal} {Phys.
  Rev. B}\ }\textbf {\bibinfo {volume} {78}},\ \bibinfo {pages} {014102}
  (\bibinfo {year} {2008})}\BibitemShut {NoStop}%
\bibitem [{\citenamefont {Chijioke}\ \emph {et~al.}(2005)\citenamefont
  {Chijioke}, \citenamefont {Nellis}, \citenamefont {Soldatov},\ and\
  \citenamefont {Silvera}}]{Chijioke2005}%
  \BibitemOpen
  \bibfield  {author} {\bibinfo {author} {\bibfnamefont {A.~D.}\ \bibnamefont
  {Chijioke}}, \bibinfo {author} {\bibfnamefont {W.~J.}\ \bibnamefont
  {Nellis}}, \bibinfo {author} {\bibfnamefont {A.}~\bibnamefont {Soldatov}}, \
  and\ \bibinfo {author} {\bibfnamefont {I.~F.}\ \bibnamefont {Silvera}},\
  }\href {\doibase 10.1063/1.2135877} {\bibfield  {journal} {\bibinfo
  {journal} {J. Appl. Phys.}\ }\textbf {\bibinfo {volume} {98}},\ \bibinfo
  {pages} {114905} (\bibinfo {year} {2005})}\BibitemShut {NoStop}%
\bibitem [{\citenamefont {Hammersley}\ \emph {et~al.}(1996)\citenamefont
  {Hammersley}, \citenamefont {Svensson}, \citenamefont {Hanfland},
  \citenamefont {Fitch},\ and\ \citenamefont {Hausermann}}]{Hammersley1996}%
  \BibitemOpen
  \bibfield  {author} {\bibinfo {author} {\bibfnamefont {A.~P.}\ \bibnamefont
  {Hammersley}}, \bibinfo {author} {\bibfnamefont {S.~O.}\ \bibnamefont
  {Svensson}}, \bibinfo {author} {\bibfnamefont {M.}~\bibnamefont {Hanfland}},
  \bibinfo {author} {\bibfnamefont {A.~N.}\ \bibnamefont {Fitch}}, \ and\
  \bibinfo {author} {\bibfnamefont {D.}~\bibnamefont {Hausermann}},\ }\href
  {\doibase 10.1080/08957959608201408} {\bibfield  {journal} {\bibinfo
  {journal} {High Press. Res.}\ }\textbf {\bibinfo {volume} {14}},\ \bibinfo
  {pages} {235} (\bibinfo {year} {1996})}\BibitemShut {NoStop}%
\bibitem [{\citenamefont {Larson}\ and\ \citenamefont {{Von
  Dreele}}(2000)}]{Larson2000}%
  \BibitemOpen
  \bibfield  {author} {\bibinfo {author} {\bibfnamefont {A.~C.}\ \bibnamefont
  {Larson}}\ and\ \bibinfo {author} {\bibfnamefont {R.~B.}\ \bibnamefont {{Von
  Dreele}}},\ }\href@noop {} {\emph {\bibinfo {title} {{General Structure
  Analysis System (GSAS)}}}},\ \bibinfo {type} {Tech. Rep.}\ (\bibinfo
  {institution} {Los Alamos National Laboratory, Report LAUR},\ \bibinfo {year}
  {2000})\BibitemShut {NoStop}%
\bibitem [{\citenamefont {Toby}(2001)}]{Toby2001}%
  \BibitemOpen
  \bibfield  {author} {\bibinfo {author} {\bibfnamefont {B.~H.}\ \bibnamefont
  {Toby}},\ }\href {\doibase 10.1107/S0021889801002242} {\bibfield  {journal}
  {\bibinfo  {journal} {J. Appl. Crystallogr.}\ }\textbf {\bibinfo {volume}
  {34}},\ \bibinfo {pages} {210} (\bibinfo {year} {2001})}\BibitemShut
  {NoStop}%
\bibitem [{\citenamefont {Dadashev}\ \emph {et~al.}(2001)\citenamefont
  {Dadashev}, \citenamefont {Pasternak}, \citenamefont {Rozenberg},\ and\
  \citenamefont {Taylor}}]{Dadashev2001}%
  \BibitemOpen
  \bibfield  {author} {\bibinfo {author} {\bibfnamefont {A.}~\bibnamefont
  {Dadashev}}, \bibinfo {author} {\bibfnamefont {M.~P.}\ \bibnamefont
  {Pasternak}}, \bibinfo {author} {\bibfnamefont {G.~K.}\ \bibnamefont
  {Rozenberg}}, \ and\ \bibinfo {author} {\bibfnamefont {R.~D.}\ \bibnamefont
  {Taylor}},\ }\href {\doibase 10.1063/1.1370561} {\bibfield  {journal}
  {\bibinfo  {journal} {Rev. Sci. Instrum.}\ }\textbf {\bibinfo {volume}
  {72}},\ \bibinfo {pages} {2633} (\bibinfo {year} {2001})}\BibitemShut
  {NoStop}%
\bibitem [{\citenamefont {Haskel}\ \emph {et~al.}(2007)\citenamefont {Haskel},
  \citenamefont {Tseng}, \citenamefont {Lang},\ and\ \citenamefont
  {Sinogeikin}}]{Haskel2007}%
  \BibitemOpen
  \bibfield  {author} {\bibinfo {author} {\bibfnamefont {D.}~\bibnamefont
  {Haskel}}, \bibinfo {author} {\bibfnamefont {Y.~C.}\ \bibnamefont {Tseng}},
  \bibinfo {author} {\bibfnamefont {J.~C.}\ \bibnamefont {Lang}}, \ and\
  \bibinfo {author} {\bibfnamefont {S.}~\bibnamefont {Sinogeikin}},\ }\href
  {\doibase 10.1063/1.2773800} {\bibfield  {journal} {\bibinfo  {journal} {Rev.
  Sci. Instrum.}\ }\textbf {\bibinfo {volume} {78}},\ \bibinfo {pages} {083904}
  (\bibinfo {year} {2007})}\BibitemShut {NoStop}%
\bibitem [{\citenamefont {Newville}(2001)}]{Newville2001}%
  \BibitemOpen
  \bibfield  {author} {\bibinfo {author} {\bibfnamefont {M.}~\bibnamefont
  {Newville}},\ }\href {\doibase 10.1107/S0909049500016964} {\bibfield
  {journal} {\bibinfo  {journal} {J. Synchrotron Radiat.}\ }\textbf {\bibinfo
  {volume} {8}},\ \bibinfo {pages} {322} (\bibinfo {year} {2001})}\BibitemShut
  {NoStop}%
\bibitem [{\citenamefont {Ravel}\ and\ \citenamefont
  {Newville}(2005)}]{Ravel2005}%
  \BibitemOpen
  \bibfield  {author} {\bibinfo {author} {\bibfnamefont {B.}~\bibnamefont
  {Ravel}}\ and\ \bibinfo {author} {\bibfnamefont {M.}~\bibnamefont
  {Newville}},\ }\href {\doibase 10.1107/S0909049505012719} {\bibfield
  {journal} {\bibinfo  {journal} {J. Synchrotron Radiat.}\ }\textbf {\bibinfo
  {volume} {12}},\ \bibinfo {pages} {537} (\bibinfo {year} {2005})}\BibitemShut
  {NoStop}%
\bibitem [{\citenamefont {Sayers}\ \emph {et~al.}(1971)\citenamefont {Sayers},
  \citenamefont {Stern},\ and\ \citenamefont {Lytle}}]{Sayers1971}%
  \BibitemOpen
  \bibfield  {author} {\bibinfo {author} {\bibfnamefont {D.}~\bibnamefont
  {Sayers}}, \bibinfo {author} {\bibfnamefont {E.}~\bibnamefont {Stern}}, \
  and\ \bibinfo {author} {\bibfnamefont {F.}~\bibnamefont {Lytle}},\ }\href
  {\doibase 10.1103/PhysRevLett.27.1204} {\bibfield  {journal} {\bibinfo
  {journal} {Phys. Rev. Lett.}\ }\textbf {\bibinfo {volume} {27}},\ \bibinfo
  {pages} {1204} (\bibinfo {year} {1971})}\BibitemShut {NoStop}%
\bibitem [{\citenamefont {Ankudinov}\ \emph {et~al.}(1998)\citenamefont
  {Ankudinov}, \citenamefont {Rehr},\ and\ \citenamefont
  {Conradson}}]{Ankudinov1998}%
  \BibitemOpen
  \bibfield  {author} {\bibinfo {author} {\bibfnamefont {A.~L.}\ \bibnamefont
  {Ankudinov}}, \bibinfo {author} {\bibfnamefont {J.~J.}\ \bibnamefont {Rehr}},
  \ and\ \bibinfo {author} {\bibfnamefont {S.~D.}\ \bibnamefont {Conradson}},\
  }\href {\doibase 10.1103/PhysRevB.58.7565} {\bibfield  {journal} {\bibinfo
  {journal} {Phys. Rev. B}\ }\textbf {\bibinfo {volume} {58}},\ \bibinfo
  {pages} {7565} (\bibinfo {year} {1998})}\BibitemShut {NoStop}%
\bibitem [{\citenamefont {Rehr}\ and\ \citenamefont {Albers}(2000)}]{Rehr2000}%
  \BibitemOpen
  \bibfield  {author} {\bibinfo {author} {\bibfnamefont {J.~J.}\ \bibnamefont
  {Rehr}}\ and\ \bibinfo {author} {\bibfnamefont {R.~C.}\ \bibnamefont
  {Albers}},\ }\href {\doibase 10.1103/RevModPhys.72.621} {\bibfield  {journal}
  {\bibinfo  {journal} {Rev. Mod. Phys.}\ }\textbf {\bibinfo {volume} {72}},\
  \bibinfo {pages} {621} (\bibinfo {year} {2000})}\BibitemShut {NoStop}%
\bibitem [{\citenamefont {Hedin}\ and\ \citenamefont
  {Lundqvist}(1970)}]{Hedin1970}%
  \BibitemOpen
  \bibfield  {author} {\bibinfo {author} {\bibfnamefont {L.}~\bibnamefont
  {Hedin}}\ and\ \bibinfo {author} {\bibfnamefont {S.}~\bibnamefont
  {Lundqvist}},\ }\href {\doibase 10.1016/S0081-1947(08)60615-3} {\bibfield
  {journal} {\bibinfo  {journal} {Solid State Phys.}\ }\textbf {\bibinfo
  {volume} {23}},\ \bibinfo {pages} {1} (\bibinfo {year} {1970})}\BibitemShut
  {NoStop}%
\bibitem [{\citenamefont {Blaha}\ \emph {et~al.}(2001)\citenamefont {Blaha},
  \citenamefont {Schwarz}, \citenamefont {Madsen}, \citenamefont {Kvasnicka},\
  and\ \citenamefont {Luitz}}]{Blaha2001}%
  \BibitemOpen
  \bibfield  {author} {\bibinfo {author} {\bibfnamefont {P.}~\bibnamefont
  {Blaha}}, \bibinfo {author} {\bibfnamefont {K.}~\bibnamefont {Schwarz}},
  \bibinfo {author} {\bibfnamefont {G.~J.~H.}\ \bibnamefont {Madsen}}, \bibinfo
  {author} {\bibfnamefont {D.}~\bibnamefont {Kvasnicka}}, \ and\ \bibinfo
  {author} {\bibfnamefont {J.}~\bibnamefont {Luitz}},\ }\href@noop {} {\emph
  {\bibinfo {title} {{WIEN2k: An Augmented Plane Wave + Local Orbitals Program
  for Calculating Crystal Properties}}}}\ (\bibinfo  {publisher} {Techn.
  Universitat},\ \bibinfo {address} {Wien, Austria},\ \bibinfo {year}
  {2001})\BibitemShut {NoStop}%
\bibitem [{\citenamefont {Perdew}\ \emph {et~al.}(1996)\citenamefont {Perdew},
  \citenamefont {Burke},\ and\ \citenamefont {Ernzerhof}}]{Perdew1996}%
  \BibitemOpen
  \bibfield  {author} {\bibinfo {author} {\bibfnamefont {J.~P.}\ \bibnamefont
  {Perdew}}, \bibinfo {author} {\bibfnamefont {K.}~\bibnamefont {Burke}}, \
  and\ \bibinfo {author} {\bibfnamefont {M.}~\bibnamefont {Ernzerhof}},\ }\href
  {\doibase 10.1103/PhysRevLett.77.3865} {\bibfield  {journal} {\bibinfo
  {journal} {Phys. Rev. Lett.}\ }\textbf {\bibinfo {volume} {77}},\ \bibinfo
  {pages} {3865} (\bibinfo {year} {1996})}\BibitemShut {NoStop}%
\bibitem [{\citenamefont {Anderson}\ and\ \citenamefont
  {Swenson}(1983)}]{Anderson1983}%
  \BibitemOpen
  \bibfield  {author} {\bibinfo {author} {\bibfnamefont {M.}~\bibnamefont
  {Anderson}}\ and\ \bibinfo {author} {\bibfnamefont {C.}~\bibnamefont
  {Swenson}},\ }\href {\doibase 10.1103/PhysRevB.28.5395} {\bibfield  {journal}
  {\bibinfo  {journal} {Phys. Rev. B}\ }\textbf {\bibinfo {volume} {28}},\
  \bibinfo {pages} {5395} (\bibinfo {year} {1983})}\BibitemShut {NoStop}%
\bibitem [{\citenamefont {Winzenick}\ \emph {et~al.}(1994)\citenamefont
  {Winzenick}, \citenamefont {Vijayakumar},\ and\ \citenamefont
  {Holzapfel}}]{Winzenick1994}%
  \BibitemOpen
  \bibfield  {author} {\bibinfo {author} {\bibfnamefont {M.}~\bibnamefont
  {Winzenick}}, \bibinfo {author} {\bibfnamefont {V.}~\bibnamefont
  {Vijayakumar}}, \ and\ \bibinfo {author} {\bibfnamefont {W.}~\bibnamefont
  {Holzapfel}},\ }\href {\doibase 10.1103/PhysRevB.50.12381} {\bibfield
  {journal} {\bibinfo  {journal} {Phys. Rev. B}\ }\textbf {\bibinfo {volume}
  {50}},\ \bibinfo {pages} {12381} (\bibinfo {year} {1994})}\BibitemShut
  {NoStop}%
\bibitem [{\citenamefont {Anderson}\ and\ \citenamefont
  {Swenson}(1985)}]{Anderson1985}%
  \BibitemOpen
  \bibfield  {author} {\bibinfo {author} {\bibfnamefont {M.}~\bibnamefont
  {Anderson}}\ and\ \bibinfo {author} {\bibfnamefont {C.}~\bibnamefont
  {Swenson}},\ }\href {\doibase 10.1103/PhysRevB.31.668} {\bibfield  {journal}
  {\bibinfo  {journal} {Phys. Rev. B}\ }\textbf {\bibinfo {volume} {31}},\
  \bibinfo {pages} {668} (\bibinfo {year} {1985})}\BibitemShut {NoStop}%
\bibitem [{\citenamefont {Birch}(1947)}]{Birch1947}%
  \BibitemOpen
  \bibfield  {author} {\bibinfo {author} {\bibfnamefont {F.}~\bibnamefont
  {Birch}},\ }\href {\doibase 10.1103/PhysRev.71.809} {\bibfield  {journal}
  {\bibinfo  {journal} {Phys. Rev.}\ }\textbf {\bibinfo {volume} {71}},\
  \bibinfo {pages} {809} (\bibinfo {year} {1947})}\BibitemShut {NoStop}%
\bibitem [{\citenamefont {McMahan}\ \emph {et~al.}(1998)\citenamefont
  {McMahan}, \citenamefont {Huscroft}, \citenamefont {Scalettar},\ and\
  \citenamefont {Pollock}}]{McMahan1998}%
  \BibitemOpen
  \bibfield  {author} {\bibinfo {author} {\bibfnamefont {A.~K.}\ \bibnamefont
  {McMahan}}, \bibinfo {author} {\bibfnamefont {C.}~\bibnamefont {Huscroft}},
  \bibinfo {author} {\bibfnamefont {R.~T.}\ \bibnamefont {Scalettar}}, \ and\
  \bibinfo {author} {\bibfnamefont {E.~L.}\ \bibnamefont {Pollock}},\ }\href
  {\doibase 10.1023/A:1008698422183} {\bibfield  {journal} {\bibinfo  {journal}
  {J. Comput. Mater. Des.}\ }\textbf {\bibinfo {volume} {5}},\ \bibinfo {pages}
  {131} (\bibinfo {year} {1998})}\BibitemShut {NoStop}%
\bibitem [{\citenamefont {Lindbaum}\ \emph {et~al.}(2003)\citenamefont
  {Lindbaum}, \citenamefont {Heathman}, \citenamefont {Bihan}, \citenamefont
  {Haire}, \citenamefont {Idiri},\ and\ \citenamefont {Lander}}]{Lindbaum2003}%
  \BibitemOpen
  \bibfield  {author} {\bibinfo {author} {\bibfnamefont {A.}~\bibnamefont
  {Lindbaum}}, \bibinfo {author} {\bibfnamefont {S.}~\bibnamefont {Heathman}},
  \bibinfo {author} {\bibfnamefont {T.~L.}\ \bibnamefont {Bihan}}, \bibinfo
  {author} {\bibfnamefont {R.~G.}\ \bibnamefont {Haire}}, \bibinfo {author}
  {\bibfnamefont {M.}~\bibnamefont {Idiri}}, \ and\ \bibinfo {author}
  {\bibfnamefont {G.~H.}\ \bibnamefont {Lander}},\ }\href {\doibase
  10.1088/0953-8984/15/28/371} {\bibfield  {journal} {\bibinfo  {journal} {J.
  Phys. Condens. Matter}\ }\textbf {\bibinfo {volume} {15}},\ \bibinfo {pages}
  {S2297} (\bibinfo {year} {2003})}\BibitemShut {NoStop}%
\bibitem [{\citenamefont {Akahama}\ \emph {et~al.}(2005)\citenamefont
  {Akahama}, \citenamefont {Fujihisa},\ and\ \citenamefont
  {Kawamura}}]{Akahama2005}%
  \BibitemOpen
  \bibfield  {author} {\bibinfo {author} {\bibfnamefont {Y.}~\bibnamefont
  {Akahama}}, \bibinfo {author} {\bibfnamefont {H.}~\bibnamefont {Fujihisa}}, \
  and\ \bibinfo {author} {\bibfnamefont {H.}~\bibnamefont {Kawamura}},\ }\href
  {\doibase 10.1103/PhysRevLett.94.195503} {\bibfield  {journal} {\bibinfo
  {journal} {Phys. Rev. Lett.}\ }\textbf {\bibinfo {volume} {94}},\ \bibinfo
  {pages} {195503} (\bibinfo {year} {2005})}\BibitemShut {NoStop}%
\bibitem [{\citenamefont {Samudrala}\ \emph {et~al.}(2012)\citenamefont
  {Samudrala}, \citenamefont {Tsoi},\ and\ \citenamefont
  {Vohra}}]{Samudrala2012}%
  \BibitemOpen
  \bibfield  {author} {\bibinfo {author} {\bibfnamefont {G.~K.}\ \bibnamefont
  {Samudrala}}, \bibinfo {author} {\bibfnamefont {G.~M.}\ \bibnamefont {Tsoi}},
  \ and\ \bibinfo {author} {\bibfnamefont {Y.~K.}\ \bibnamefont {Vohra}},\
  }\href {\doibase 10.1088/0953-8984/24/36/362201} {\bibfield  {journal}
  {\bibinfo  {journal} {J. Phys. Condens. Matter}\ }\textbf {\bibinfo {volume}
  {24}},\ \bibinfo {pages} {362201} (\bibinfo {year} {2012})}\BibitemShut
  {NoStop}%
\bibitem [{\citenamefont {Fabbris}\ \emph {et~al.}(2013)\citenamefont
  {Fabbris}, \citenamefont {Matsuoka}, \citenamefont {Lim}, \citenamefont
  {Mardegan}, \citenamefont {Shimizu}, \citenamefont {Haskel},\ and\
  \citenamefont {Schilling}}]{Fabbris2013}%
  \BibitemOpen
  \bibfield  {author} {\bibinfo {author} {\bibfnamefont {G.}~\bibnamefont
  {Fabbris}}, \bibinfo {author} {\bibfnamefont {T.}~\bibnamefont {Matsuoka}},
  \bibinfo {author} {\bibfnamefont {J.}~\bibnamefont {Lim}}, \bibinfo {author}
  {\bibfnamefont {J.}~\bibnamefont {Mardegan}}, \bibinfo {author}
  {\bibfnamefont {K.}~\bibnamefont {Shimizu}}, \bibinfo {author} {\bibfnamefont
  {D.}~\bibnamefont {Haskel}}, \ and\ \bibinfo {author} {\bibfnamefont
  {J.}~\bibnamefont {Schilling}},\ }\href {\doibase 10.1103/PhysRevB.88.245103}
  {\bibfield  {journal} {\bibinfo  {journal} {Phys. Rev. B}\ }\textbf {\bibinfo
  {volume} {88}},\ \bibinfo {pages} {245103} (\bibinfo {year}
  {2013})}\BibitemShut {NoStop}%
\bibitem [{\citenamefont {Johansson}(1974)}]{Johansson1974}%
  \BibitemOpen
  \bibfield  {author} {\bibinfo {author} {\bibfnamefont {B.}~\bibnamefont
  {Johansson}},\ }\href {\doibase 10.1080/14786439808206574} {\bibfield
  {journal} {\bibinfo  {journal} {Philos. Mag.}\ }\textbf {\bibinfo {volume}
  {30}},\ \bibinfo {pages} {469} (\bibinfo {year} {1974})}\BibitemShut
  {NoStop}%
\bibitem [{\citenamefont {Allen}\ and\ \citenamefont
  {Martin}(1982)}]{Allen1982}%
  \BibitemOpen
  \bibfield  {author} {\bibinfo {author} {\bibfnamefont {J.}~\bibnamefont
  {Allen}}\ and\ \bibinfo {author} {\bibfnamefont {R.}~\bibnamefont {Martin}},\
  }\href {\doibase 10.1103/PhysRevLett.49.1106} {\bibfield  {journal} {\bibinfo
   {journal} {Phys. Rev. Lett.}\ }\textbf {\bibinfo {volume} {49}},\ \bibinfo
  {pages} {1106} (\bibinfo {year} {1982})}\BibitemShut {NoStop}%
\bibitem [{\citenamefont {Rueff}\ \emph {et~al.}(2006)\citenamefont {Rueff},
  \citenamefont {Iti\'{e}}, \citenamefont {Taguchi}, \citenamefont {Hague},
  \citenamefont {Mariot}, \citenamefont {Delaunay}, \citenamefont {Kappler},\
  and\ \citenamefont {Jaouen}}]{Rueff2006}%
  \BibitemOpen
  \bibfield  {author} {\bibinfo {author} {\bibfnamefont {J.-P.}\ \bibnamefont
  {Rueff}}, \bibinfo {author} {\bibfnamefont {J.-P.}\ \bibnamefont {Iti\'{e}}},
  \bibinfo {author} {\bibfnamefont {M.}~\bibnamefont {Taguchi}}, \bibinfo
  {author} {\bibfnamefont {C.}~\bibnamefont {Hague}}, \bibinfo {author}
  {\bibfnamefont {J.-M.}\ \bibnamefont {Mariot}}, \bibinfo {author}
  {\bibfnamefont {R.}~\bibnamefont {Delaunay}}, \bibinfo {author}
  {\bibfnamefont {J.-P.}\ \bibnamefont {Kappler}}, \ and\ \bibinfo {author}
  {\bibfnamefont {N.}~\bibnamefont {Jaouen}},\ }\href {\doibase
  10.1103/PhysRevLett.96.237403} {\bibfield  {journal} {\bibinfo  {journal}
  {Phys. Rev. Lett.}\ }\textbf {\bibinfo {volume} {96}},\ \bibinfo {pages}
  {237403} (\bibinfo {year} {2006})}\BibitemShut {NoStop}%
\bibitem [{\citenamefont {Bradley}\ \emph {et~al.}(2012)\citenamefont
  {Bradley}, \citenamefont {Moore}, \citenamefont {Lipp}, \citenamefont
  {Mattern}, \citenamefont {Pacold}, \citenamefont {Seidler}, \citenamefont
  {Chow}, \citenamefont {Rod}, \citenamefont {Xiao},\ and\ \citenamefont
  {Evans}}]{Bradley2012}%
  \BibitemOpen
  \bibfield  {author} {\bibinfo {author} {\bibfnamefont {J.}~\bibnamefont
  {Bradley}}, \bibinfo {author} {\bibfnamefont {K.}~\bibnamefont {Moore}},
  \bibinfo {author} {\bibfnamefont {M.}~\bibnamefont {Lipp}}, \bibinfo {author}
  {\bibfnamefont {B.}~\bibnamefont {Mattern}}, \bibinfo {author} {\bibfnamefont
  {J.}~\bibnamefont {Pacold}}, \bibinfo {author} {\bibfnamefont
  {G.}~\bibnamefont {Seidler}}, \bibinfo {author} {\bibfnamefont
  {P.}~\bibnamefont {Chow}}, \bibinfo {author} {\bibfnamefont {E.}~\bibnamefont
  {Rod}}, \bibinfo {author} {\bibfnamefont {Y.}~\bibnamefont {Xiao}}, \ and\
  \bibinfo {author} {\bibfnamefont {W.}~\bibnamefont {Evans}},\ }\href
  {\doibase 10.1103/PhysRevB.85.100102} {\bibfield  {journal} {\bibinfo
  {journal} {Phys. Rev. B}\ }\textbf {\bibinfo {volume} {85}},\ \bibinfo
  {pages} {100102} (\bibinfo {year} {2012})}\BibitemShut {NoStop}%
\bibitem [{Lip(2012)}]{Lipp2012}%
  \BibitemOpen
  \href {\doibase 10.1103/PhysRevLett.109.195705} {\bibfield  {journal}
  {\bibinfo  {journal} {Phys. Rev. Lett.}\ }\textbf {\bibinfo {volume} {109}},\
  \bibinfo {pages} {195705} (\bibinfo {year} {2012})}\BibitemShut {NoStop}%
\bibitem [{\citenamefont {Johansson}\ \emph {et~al.}(2014)\citenamefont
  {Johansson}, \citenamefont {Luo}, \citenamefont {Li},\ and\ \citenamefont
  {Ahuja}}]{Johansson2014}%
  \BibitemOpen
  \bibfield  {author} {\bibinfo {author} {\bibfnamefont {B.}~\bibnamefont
  {Johansson}}, \bibinfo {author} {\bibfnamefont {W.}~\bibnamefont {Luo}},
  \bibinfo {author} {\bibfnamefont {S.}~\bibnamefont {Li}}, \ and\ \bibinfo
  {author} {\bibfnamefont {R.}~\bibnamefont {Ahuja}},\ }\href {\doibase
  10.1038/srep06398} {\bibfield  {journal} {\bibinfo  {journal} {Sci. Rep.}\
  }\textbf {\bibinfo {volume} {4}},\ \bibinfo {pages} {6398} (\bibinfo {year}
  {2014})}\BibitemShut {NoStop}%
\bibitem [{\citenamefont {Mattheiss}(1964)}]{Mattheiss1964}%
  \BibitemOpen
  \bibfield  {author} {\bibinfo {author} {\bibfnamefont {L.}~\bibnamefont
  {Mattheiss}},\ }\href {\doibase 10.1103/PhysRev.133.A1399} {\bibfield
  {journal} {\bibinfo  {journal} {Phys. Rev.}\ }\textbf {\bibinfo {volume}
  {133}},\ \bibinfo {pages} {A1399} (\bibinfo {year} {1964})}\BibitemShut
  {NoStop}%
\bibitem [{\citenamefont {Sternheimer}(1950)}]{Sternheimer1950}%
  \BibitemOpen
  \bibfield  {author} {\bibinfo {author} {\bibfnamefont {R.}~\bibnamefont
  {Sternheimer}},\ }\href {\doibase 10.1103/PhysRev.78.235} {\bibfield
  {journal} {\bibinfo  {journal} {Phys. Rev.}\ }\textbf {\bibinfo {volume}
  {78}},\ \bibinfo {pages} {235} (\bibinfo {year} {1950})}\BibitemShut
  {NoStop}%
\bibitem [{\citenamefont {Duthie}\ and\ \citenamefont
  {Pettifor}(1977)}]{Duthie1977}%
  \BibitemOpen
  \bibfield  {author} {\bibinfo {author} {\bibfnamefont {J.}~\bibnamefont
  {Duthie}}\ and\ \bibinfo {author} {\bibfnamefont {D.}~\bibnamefont
  {Pettifor}},\ }\href {\doibase 10.1103/PhysRevLett.38.564} {\bibfield
  {journal} {\bibinfo  {journal} {Phys. Rev. Lett.}\ }\textbf {\bibinfo
  {volume} {38}},\ \bibinfo {pages} {564} (\bibinfo {year} {1977})}\BibitemShut
  {NoStop}%
\end{thebibliography}%

\end{document}